\documentclass{article}

\usepackage{microtype}
\usepackage{graphicx}
\usepackage{subcaption}
\usepackage{booktabs}
\usepackage{url}
\usepackage{hyperref}

\usepackage{pifont}
\usepackage{fontawesome}
\usepackage{natbib}

\usepackage[accepted]{icml2026}

\usepackage{amsmath,amssymb,mathtools,amsthm}
\newcommand{\cmark}{\ding{51}} 
\newcommand{\xmark}{\ding{55}} 
\usepackage[capitalize,noabbrev]{cleveref}

\theoremstyle{plain}

\theoremstyle{definition}

\theoremstyle{remark}

\icmltitlerunning{From Lossy to Verified: A Provenance-Aware Tiered Memory for Agents}

\usepackage{listings}
\usepackage{xcolor}
\usepackage{verbatim}
\usepackage{enumitem}
\usepackage[most]{tcolorbox} 
\usepackage{float}
\usepackage{xspace}

\tcbset{
  aibox/.style={
    width=474.18663pt,
    top=10pt,
    colback=white,
    colframe=black,
    colbacktitle=black,
    enhanced,
    center,
    attach boxed title to top left={yshift=-0.1in,xshift=0.15in},
    boxed title style={boxrule=0pt,colframe=white,},
  }
}
\newtcolorbox{AIbox}[2][]{aibox,title=#2,#1}
\newcommand{\Msum}{\mathcal{M}_S}        
\newcommand{\Mraw}{\mathcal{M}_R}        

\makeatletter
\renewcommand{\ICML@appearing}{Preprint.}
\makeatother
\begin{document}

\twocolumn[
\icmltitle{From Lossy to Verified: A Provenance-Aware Tiered Memory for Agents}

\begin{icmlauthorlist}
\icmlauthor{Qiming Zhu}{sds}
\icmlauthor{Shunian Chen}{sds}
\icmlauthor{Rui Yu}{sds}
\icmlauthor{Zhehao Wu}{sds}
\icmlauthor{Benyou Wang}{sds}
\end{icmlauthorlist}

\icmlaffiliation{sds}{School of Data Science, The Chinese University of Hong Kong, Shenzhen}

\icmlcorrespondingauthor{Qiming Zhu}{qimingzhu@link.cuhk.edu.cn} 

\icmlkeywords{LLM Agents, Agent Memory, Hierarchical Memory, Routing, GRPO, Cost-Aware Optimization}

\vskip 0.3in
]
\printAffiliationsAndNotice{}

\begin{abstract}
Long-horizon agents often compress interaction histories into write-time summaries. This creates a fundamental \textbf{write-before-query} barrier: compression decisions are made before the system knows what a future query will hinge on. As a result, summaries can cause \textbf{unverifiable omissions}---decisive constraints (e.g., allergies) may be dropped, leaving the agent unable to justify an answer with traceable evidence. Retaining raw logs restores an authoritative source of truth, but grounding on raw logs by default is expensive: many queries are answerable from summaries, yet raw grounding still requires processing far longer contexts, inflating token consumption and latency.

We propose \textbf{TierMem}, a provenance-linked framework that casts retrieval as an inference-time evidence allocation problem. TierMem uses a \textbf{two-tier} memory hierarchy to answer with the \emph{cheapest sufficient evidence}: it queries a fast summary index by default, and a runtime \textbf{sufficiency router} \textsc{Escalates} to an immutable raw-log store only when summary evidence is insufficient. TierMem then writes back verified findings as new summary units linked to their raw sources. On LoCoMo, TierMem achieves \textbf{0.851} accuracy (vs.\ \textbf{0.873} raw-only) while reducing input tokens by \textbf{54.1\%} and latency by \textbf{60.7\%}. Code is available at \url{https://github.com/FreedomIntelligence/Tiermem}
\end{abstract}

\section{Introduction}
\label{sec:intro}
Long-horizon language agents increasingly need to answer over interaction histories spanning weeks or months \cite{chen_extending_2023,liu_lost_2023}. 
At scale, their memory must be simultaneously \emph{evidence-faithful} (grounded in source), \emph{auditable} (traceable to provenance), and \emph{economical} (low latency and cost). 
Yet most current systems rely on \emph{write-time compression} into summaries or extracted facts \cite{chhikara_mem0_2025,fang_lightmem_2025}, committing to what to retain \emph{before} the agent knows what future queries will require.

\paragraph{Summaries at write-time may erase decisive evidence.}
Write-time compression forces irreversible retention decisions under uncertainty, requiring a preemptive bet on saliency before the query distribution is known \cite{zheng2025goaldirected}.
Consequently, even state-of-the-art summarizers discard nuances that later prove decisive \cite{kang2025acon}.
For instance, compressing a ``severe peanut allergy'' into generic ``dietary preferences'' creates an \textbf{unverifiable omission}: when later asked ``Is this snack safe?'', the agent cannot trace or cite the specific constraint from its memory.
Thus, any fixed-budget summary admits a worst-case query it cannot support with traceable evidence. These failures are not merely occasional summarization errors; they are structural risks of lossy, write-before-query compression.

\paragraph{Raw grounding is auditable but often wasteful.}
A natural response is to avoid compression and always ground on raw logs~\citep{asai_self-rag_2023,yan_general_2025,du_memr3_2025}.
This improves auditability, but it is often economically and algorithmically inefficient.
Naively feeding long histories or globally retrieving large raw contexts incurs substantial token and latency costs, and long inputs can even degrade effective context utilization due to position-sensitive failures and attention dilution~\citep{liu_lost_2023,zhang_recursive_2025}.
Crucially, most queries do \emph{not} require raw-level evidence.

\paragraph{Bridging the gap: Answer with the cheapest sufficient evidence.}
These two extremes suggest that long-horizon memory access should be \emph{query-conditioned}.
Given a query, the agent should retrieve the \emph{lowest-cost} memory granularity that still provides \emph{sufficient evidence} for faithful, auditable answering. This principle echoes recent adaptive RAG work that selectively accesses external evidence to trade off quality against retrieval cost \cite{su2025fastorbetter,wang2025targ}.
We view this as an \textbf{inference-time evidence allocation problem}: decide at runtime when cheap summaries are enough and when the system must consult authoritative raw sources.
A practical system must therefore (i) detect when retrieved summaries are evidence-insufficient, (ii) fall back to an auditable source of truth only in those cases, and (iii) consolidate verified findings back into cheap memory to amortize future queries.

\begin{figure*}[t]
\centering
\includegraphics[width=\textwidth]{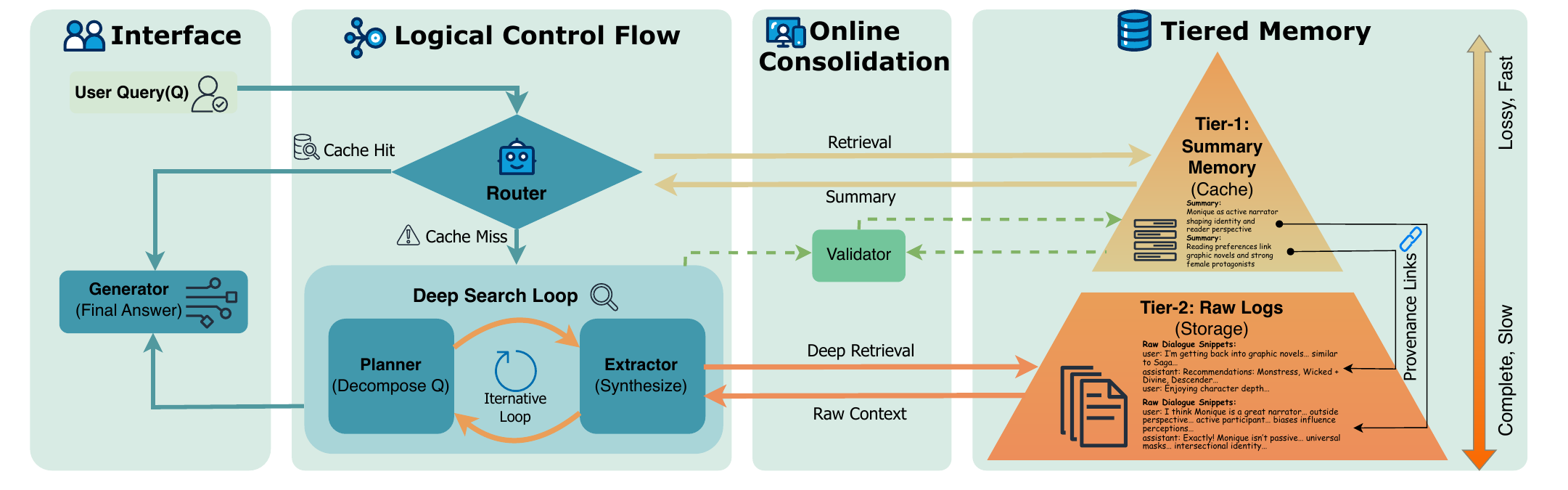}
\caption{\textbf{TierMem overview.} TierMem maintains a provenance-linked two-tier memory hierarchy.
Tier-1 is a fast summary index whose entries store (i) compact summaries and (ii) provenance links $\rho$ to supporting Tier-2 raw pages in an immutable paged log.
Given a query, TierMem retrieves Tier-1 evidence and uses a lightweight router $\pi_\theta$ to choose \textsc{Answer} (summaries only) or \textsc{Escalate} (consult linked raw pages, then run a bounded retrieval procedure if needed).
After escalation, TierMem can optionally perform verified write-back to update Tier-1 with evidence-backed findings linked to their raw sources.}
\label{fig:TierMem}
\end{figure*}

\paragraph{TierMem: linked summaries with selective escalation.}
Drawing inspiration from tiered caches, we propose \textbf{TierMem}, a linked two-tier memory hierarchy governed by a runtime sufficiency check.
As shown in Figure~\ref{fig:TierMem}, Tier-1 is a fast summary index whose units are \emph{explicitly linked} to their source raw pages via provenance pointers, enabling targeted escalation.
Tier-2 is an immutable raw-log store that serves as an auditable source of truth.
Given a user query, TierMem first retrieves from Tier-1 and applies a lightweight \textbf{miss detector} (router) that judges whether the retrieved summary evidence is sufficient.
On a \textsc{hit}, the agent answers immediately from summaries.
On a \textsc{miss}, the system \textsc{Escalates} to the linked raw pages (and, if needed, bounded global retrieval) for deep grounding.
After escalation, TierMem performs \textbf{verified write-back}: it distills only \emph{grounded} evidence into new Tier-1 units linked back to supporting raw pages, enabling \emph{online consolidation} without sacrificing auditability.

Across two long-horizon conversational memory benchmarks, \textbf{LoCoMo} and \textbf{LongMemEval}, TierMem nearly matches raw-only accuracy while substantially reducing query-time cost by escalating only when necessary.
On \textbf{LoCoMo}, TierMem reaches \textbf{0.851} accuracy versus \textbf{0.873} for always-raw, while reducing average input tokens by \textbf{54.1\%} and latency by \textbf{60.7\%} (3{,}396 vs.\ 7{,}398 tokens; 6.76s vs.\ 17.18s).
On \textbf{LongMemEval}, where summary-only methods degrade sharply under long-range updates, TierMem mitigates compression loss by routing evidence-insufficient cases to raw grounding.

\paragraph{Contributions.}

\begin{enumerate}
    \item \textbf{Problem formulation.} We identify the \emph{write-before-query barrier} in summary-centric agent memory and cast long-horizon memory access as \emph{inference-time evidence allocation}: choose the minimal granularity that provides sufficient evidence for faithful, auditable answering.
    \item \textbf{Framework.} We introduce \textbf{TierMem}, a linked two-tier memory hierarchy with (i) \emph{provenance-linked summaries}, (ii) a lightweight \emph{miss detector} that routes \textsc{Answer} vs.\ \textsc{Escalate}, and (iii) \emph{verified write-back} for online consolidation.
    \item \textbf{Empirics.} We show TierMem improves the accuracy--efficiency trade-off on long-horizon benchmarks, recovering much of the benefit of raw grounding at substantially lower token and latency cost.
\end{enumerate}

\section{The TierMem Framework}
\label{sec:framework}

This section introduces \textbf{TierMem}, a two-tier memory framework that casts long-horizon retrieval as an \emph{inference-time evidence allocation} problem.
TierMem prioritizes efficiency by defaulting to a fast summary tier, yet eliminates unverifiable omissions by \textsc{Escalating} to immutable raw logs when summaries lack sufficient detail.
Crucially, the system employs a \emph{provenance-linked} architecture to ensure that all consolidated memories remain traceable to their source.Detailed architectural design and implementation specifics are provided in Appendix~\ref{app:impl_details}.

\subsection{Architecture: Provenance-Linked 2 Tier Memory}
\label{sec:arch}

TierMem organizes memory into two distinct levels of granularity, bridged by \textbf{explicit provenance links} to ensure traceability.

\vspace{0.25em}
\noindent\textbf{Tier-2: Immutable Paged Raw Log ($\Mraw$).}
To guarantee auditability, Tier-2 stores the full interaction history as immutable, fixed-size pages (e.g., $\sim$1000 tokens), each assigned a stable identifier.
Unlike sliding windows, paged storage ensures that evidence units remain static, allowing stable citation even as the context grows.

\vspace{0.25em}
\noindent\textbf{Tier-1: Provenance Index ($\Msum$).}
Tier-1 acts as a semantic cache for fast retrieval.
Each entry contains: (i) a concise summary, (ii) a dense embedding for retrieval, and (iii) \textbf{provenance links} ($\rho$) pointing to the specific Tier-2 page identifiers that support the summary.
These links prevent hallucinations by anchoring summaries to raw data and provide a high-recall ``warm start'' pointer during escalation.

\vspace{0.25em}
\noindent\textbf{Construction.}
The tiers are built synchronously: whenever a raw page is sealed in Tier-2, TierMem synthesizes corresponding summary entries in Tier-1, automatically recording the links $\rho$. This ensures no summary exists without a traceable raw source.

\subsection{Inference Protocol: Adaptive Routing and Consolidation}
\label{sec:protocol}

Given a query, TierMem executes a dynamic protocol (Figure~\ref{fig:TierMem}) to select the minimal sufficient evidence granularity.

\paragraph{Router Policy ($\pi_\theta$).}
The core decision maker is a lightweight policy $\pi_\theta$ that judges \emph{evidence sufficiency}.
It takes the query and retrieved Tier-1 summaries as input and outputs a binary decision: \textsc{Answer} or \textsc{Escalate}.
The policy is trained to define ``sufficiency'' strictly: it must escalate if the summary lacks precise constraints (e.g., negations, exact values, attribution) required for a faithful answer (training details in \S\ref{sec:optimization}).

\paragraph{Phase 1: Fast Path (Default).}
TierMem first retrieves top-$k$ summaries from Tier-1.
If $\pi_\theta$ outputs \textsc{Answer}, the system generates the response immediately using only these summaries.
This path minimizes latency and cost for the majority of queries where high-level context suffices.

\paragraph{Phase 2: Provenance-Guided Escalation.}
If $\pi_\theta$ outputs \textsc{Escalate}, TierMem triggers a \textbf{bounded retrieval loop} to recover missing details.
Crucially, this search is \emph{provenance-guided}: it prioritizes reading the Tier-2 pages explicitly linked ($\rho$) by the retrieved summaries, as these likely contain the omitted nuances.
If this initial look-up remains insufficient, the system performs a limited multi-hop search over Tier-2 (up to $T_{max}$ steps) to locate disconnected evidence.
This ensures deep grounding without the prohibitive cost of reading the full raw history.

\paragraph{Phase 3: Verified Write-Back.}
To prevent recurring escalations for the same topic, TierMem performs \textbf{online consolidation}.
After a successful escalation, the system extracts the newly verified details and updates Tier-1 via a \emph{retrieve-and-edit} operation: new findings are either added as new entries or merged into existing ones.
Importantly, these updates preserve the provenance links to the Tier-2 evidence used during escalation, maintaining the chain of custody.

\section{Router Optimization}
\label{sec:optimization}

We optimize the router $\pi_\theta$ to minimize inference costs while maintaining the accuracy of a raw-grounded upper bound.The full experimental setup for router training is detailed in Appendix~\ref{app:router_full}.

\subsection{Training Signal: Summary Sufficiency Labels}
\label{sec:opt_labels}

We supervise $\pi_\theta$ using a \textbf{hindsight labeling} strategy.
For a given query, we generate two answers: (1) a \emph{Summary-Only} answer, and (2) a \emph{Raw-Grounded} answer (derived from the escalation path).
An LLM judge compares these against the query requirements:
\begin{itemize}
    \item Label \textsc{Answer} (0): If the summary-only answer is correct and complete.
    \item Label \textsc{Escalate} (1): If the summary answer is vague or incorrect, but the raw-grounded answer recovers the necessary details.
\end{itemize}
This filters out ``impossible" queries (where even raw logs fail) and isolates cases where write-time compression caused an unverifiable omission.

\subsection{Optimization: Distillation and Alignment}
\label{sec:opt_train}

We employ a two-stage training process to refine the decision boundary.

\vspace{0.25em}
\noindent\textbf{Stage 1: Distillation (SFT).}
We distill routing behavior from a stronger teacher (\textbf{GPT-5}) that outputs a short rationale and the final decision. We fine-tune the router to imitate the teacher outputs, keeping only examples where the teacher decision matches our oracle label (details in Appendix~\ref{app:router_sft}).

\vspace{0.25em}
\noindent\textbf{Stage 2: Cost-Aware Alignment (GRPO).}
To explicitly manage the accuracy-efficiency trade-off, we further optimize $\pi_\theta$ using Group Relative Policy Optimization (GRPO). The reward function is defined as:
\begin{equation*} R(\tau) = R_{\mathrm{acc}}(\tau)\;-\;\lambda_{\mathrm{cost}} R_{\mathrm{cost}}(\tau)\;-\;\lambda_{\mathrm{waste}} R_{\mathrm{waste}}(\tau) \label{eq:reward} \end{equation*} 
where $R_{\mathrm{acc}}$ rewards answer correctness, $\lambda_{\mathrm{cost}}$ penalizes the computational cost of escalation, and $\lambda_{\mathrm{waste}}$ specifically penalizes ``false alarms" (escalating when summaries were actually sufficient).
This objective forces the model to escalate only when the expected gain in faithfulness outweighs the retrieval cost.
\section{Experiments}
\label{sec:experiments}

We reframe memory retrieval as a problem of \textbf{inference-time evidence allocation}.
Our central hypothesis is that aligning retrieval granularity with query complexity—defaulting to efficient summaries and \textsc{Escalating} to raw logs only when strictly necessary—can break the static trade-off between cost and faithfulness.

We evaluate whether TierMem successfully navigates this optimization landscape through four Research Questions (RQs):

\begin{description}[style=unboxed, leftmargin=0cm, font=\bfseries]
    \item[RQ1 (Trade-off):] Does TierMem improve the accuracy--efficiency trade-off compared to static summary-only and raw-only policies?

    \item[RQ2 (Control Mechanism):] Can a lightweight router reliably detect evidence insufficiency (distinguishing \textsc{Answer} vs.\ \textsc{Escalate}) with negligible inference overhead?
    \item[RQ3 (Structural Utility):] Do explicit provenance pointers (Tier-1$\rightarrow$Tier-2) measurably improve the grounding quality of escalated queries?
    \item[RQ4 (Online Amortization):] Does the consolidation of verified findings reduce future cache misses, effectively amortizing raw-access costs over time?
\end{description}

\subsection{Experimental Setup}
\label{sec:exp_setup}

\subsubsection{Baselines}
\label{sec:baselines}
We compare against representative systems spanning summary-centric and raw/controller-centric paradigms.
On the summary-centric side, \textbf{Mem0}~\cite{chhikara_mem0_2025} \textbf{LightMem}~\cite{fang_lightmem_2025}, \textbf{O-mem}~\cite{wang2025omemomnimemorypersonalized}, and \textbf{MemOS} ~\cite{li_memos_2025} primarily rely on compact written memories for fast retrieval.
On the raw/controller-centric side, \textbf{MemR3}~\cite{du_memr3_2025} and \textbf{GAM}~\cite{yan_general_2025} allocate more inference-time budget to access fine-grained evidence and improve grounding.

For TierMem, we report:
\textbf{Ours (summary)} (Tier-1 only),
\textbf{Ours (raw-data)} (always Tier-2),
\textbf{Ours (router)} (learned routing),
and \textbf{Ours (router, GPT-4.1-mini)} (a stronger LLM routing reference). To ensure a fair and consistent comparison, we \textbf{disable online write-back} for all methods in the main benchmark evaluation. We use GPT-4.1-mini as the backbone in memory systems.

\paragraph{Fair comparison within TierMem.}
To isolate \emph{memory access policy} from model choice, all TierMem variants use the same generator, the same embedding model for Tier-1 dense retrieval, and the same reranker.
Among TierMem variants, the \textbf{only} differences are (i) whether Tier-2 is consulted, and (ii) how/when escalation is triggered (always vs.\ routed).
We report efficiency measured in the QA phase (retrieval + generation) and explicitly account for routing overhead when present.

\subsubsection{Metrics and Cost Accounting}
\label{sec:metrics}
We report both \textbf{answer quality} and \textbf{query-time efficiency}, together with \textbf{routing behavior} and a specific metric for compression loss.

\paragraph{Answer quality \& Compression Loss.}
We report:
(i) \textbf{LLM-judge accuracy} (primary): fraction of queries judged correct by the benchmark protocol.
(ii) \textbf{Unverifiable Omission Rate (UOR)}: To quantify the write-before-query barrier, we measure the fraction of queries where summary-based methods fail specifically due to missing evidence (i.e., the answer can be found in our raw-data system but was lost in summaries), distinguishing these from reasoning failures.

\paragraph{Query-time efficiency (QA phase).}
We measure:
\textbf{Avg. input tokens / query} (Tok$_{\text{in}}$/Q),
\textbf{Avg. output tokens / query} (Tok$_{\text{out}}$/Q),
and \textbf{Avg. latency / query} (Lat./Q), when measured.
Unless otherwise stated, escalation uses a bounded Deep Search Loop capped at $T_{\max}{=}3$ iterations.

\paragraph{Routing behavior.}
We measure:
(i) \textbf{R-rate} (fraction of queries routed to \textsc{Escalate}),
(ii) \textbf{Recall on hard cases} (fraction of oracle-hard queries correctly escalated),
and (iii) \textbf{Router overhead} (tokens consumed by the routing step itself).

\paragraph{Token accounting for routed systems.}
For routed variants, we decompose query-time token usage into:
$
\text{Tok}_{\text{in}} = \text{Tok}_{\text{QA}} + \text{Tok}_{\text{router}}\quad\text{Tok}_{\text{out}} = \text{Tok}_{\text{gen}} + \text{Tok}_{\text{router-out}}$
and report both the total and the components (where applicable).
This makes the cost--fidelity trade-off explicit: routing should reduce $\text{Tok}_{\text{QA}}$ by avoiding unnecessary raw reads, while keeping $\text{Tok}_{\text{router}}$ small.

\subsection{Main Results: Accuracy--Efficiency trade-off}
\label{sec:main_results}

We analyze the accuracy--efficiency trade-off (RQ1) across both benchmarks.
Broadly, \textbf{Ours (summary)} is the most efficient but suffers from compression loss (the \textit{write-before-query barrier}), while \textbf{Ours (raw-data)} is the most accurate but costly.
\textbf{Ours (router)} balances this via \textbf{inference-time evidence allocation}, approaching raw-only accuracy with summary-like efficiency.

\paragraph{LoCoMo.}
The write-before-query barrier is quantifiable via UOR analysis (Table~\ref{tab:locomo_main}): summary-centric baselines exhibit high omission rates (14.7\%--23.3\%), confirming that errors often stem from evidence unavailability.
\textbf{Ours (router)} overcomes this, achieving \textbf{0.851} overall accuracy (approaching the \textbf{0.873} raw-only upper bound) while reducing average input tokens by \textbf{54.1\%} and latency by \textbf{60.7\%} compared to the raw baseline.
In contrast, \textbf{Ours (summary)} drops to 0.755 accuracy, consistent with its 15.4\% UOR.
The router provides the largest gains on queries requiring precise details (e.g., entities, temporal qualifiers) that static summaries often drop.

\paragraph{LongMemEval.}
As shown in Table~\ref{tab:longmemeval_full_results}, the summary-raw gap widens due to long-range dependencies.
Static summaries fail to retain answerable evidence for nearly 30\% of queries (e.g., 29.6\% UOR for Mem0).
\textbf{Ours (router)} effectively bridges this gap by selectively escalating hard cases, offering a superior trade-off compared to static memory formats.

\begin{table*}[t]
\centering
\small
\setlength{\tabcolsep}{3pt}
\caption{LoCoMo test results. \textbf{UOR (Unverifiable Omission Rate)} quantifies errors caused by missing evidence in summaries (lower is better). Bold indicates the best result in each column.}
\resizebox{\textwidth}{!}{
\begin{tabular}{lrrrrrrcccc}
\toprule
& \multicolumn{5}{c}{\textbf{Accuracy (LLM-Judge)} $\uparrow$} & \multicolumn{1}{c}{\textbf{UOR \%}$\downarrow$} & \multicolumn{1}{c}{\textbf{F1}$\uparrow$} & \multicolumn{3}{c}{\textbf{Efficiency}} \\
\cmidrule(lr){2-6} \cmidrule(lr){7-7} \cmidrule(lr){8-8} \cmidrule(lr){9-11}
\textbf{System} &
\multicolumn{1}{c}{\textbf{Multi}} &
\multicolumn{1}{c}{\textbf{Temp.}} &
\multicolumn{1}{c}{\textbf{Open}} &
\multicolumn{1}{c}{\textbf{Single}} &
\multicolumn{1}{c}{\textbf{Overall}} &
\multicolumn{1}{c}{\textbf{Rate}} &
\multicolumn{1}{c}{\textbf{Overall}} &
\textbf{Tok$_{\text{in}}$} $\downarrow$ &
\textbf{Tok$_{\text{out}}$} $\downarrow$ &
\textbf{Lat.(s)} $\downarrow$ \\
\midrule
Mem0~\cite{chhikara_mem0_2025} & 0.628 & 0.707 & 0.531 & 0.712 & 0.684 & 23.3 & 0.426 & 1085.9 & 8.0 & 20.32 \\
LightMem~\cite{fang_lightmem_2025} & 0.709 & 0.807 & 0.542 & 0.810 & 0.773 & 14.7 & 0.495 & 1159.5 & 6.9 & 27.69 \\
O-mem~\cite{wang2025omemomnimemorypersonalized} & 0.706 & 0.776 & 0.604 & 0.797 & 0.764 & 16.3 & 0.512 & 2178.1 & 544.2 & 1.16 \\
MemOS ~\cite{li_memos_2025} & 0.773 & 0.639 & 0.625 & 0.804 & 0.753 & 17.1 & 0.400 & 2692.3 & 13.3 & 10.80
\\
MemR3~\cite{du_memr3_2025} & 0.812 & 0.829 & 0.729 & 0.902 & 0.860 & - & 0.586 & 4585.7 & 384.1 & 12.30 \\
GAM~\cite{yan_general_2025} & \textbf{0.858} & 0.860 & \textbf{0.781} & 0.886 & 0.869 & - & 0.574 & 15773.7 & 1231.5 & 21.85 \\
\midrule
Ours (summary) & 0.656 & 0.717 & 0.667 & 0.812 & 0.755 & 15.4 & 0.514 & \textbf{260.1} & \textbf{6.1} & \textbf{1.24} \\
Ours (raw-data) & 0.812 & \textbf{0.863} & 0.719 & \textbf{0.916} & \textbf{0.873} & - & \textbf{0.604} & 7398.4 & 521.3 & 17.18 \\
\textbf{Ours (router)} & 0.794 & 0.829 & 0.708 & 0.895 & 0.851 & - & 0.581 & (3396+584) & (247+94) & 6.76 \\
Ours (router, 4.1-mini) & 0.755 & 0.850 & 0.677 & 0.904 & 0.851 & - & 0.591 & 3620.9 & 235.4 & 6.40 \\
\bottomrule
\end{tabular}
}
\label{tab:locomo_main}
\end{table*}

\begin{table*}[t]
\centering
\small
\setlength{\tabcolsep}{3pt}
\caption{Main results on LongMemEval. Summary-based methods show high UOR, indicating frequent loss of key evidence during compression.}
\resizebox{\textwidth}{!}{
\begin{tabular}{lccccccccccc}
\toprule
\textbf{Method} &
\textbf{SS-User} &
\textbf{SS-Asst} &
\textbf{SS-Pref} &
\textbf{Multi-S} &
\textbf{Know. Upd} &
\textbf{Temp. Reas} &
\textbf{Overall} &
\textbf{UOR \%}$\downarrow$ &
\textbf{Tok$_{\text{in}}$/Q} &
\textbf{Tok$_{\text{out}}$/Q} &
\textbf{Lat./Q} \\
\midrule
Mem0 & 0.800 & 0.589 & 0.533 & 0.571 & 0.667 & 0.489 & 0.596 & 29.6 & 1268 & 9 & 15.2 \\
LightMem & 0.871 & 0.339 & 0.767 & \textbf{0.790} & \textbf{0.885} & \textbf{0.700} & 0.716 & 16.8 & 1160 & 7 & 7.87 \\
O-mem & 0.957 & 0.446 & 0.773 & 0.632 & 0.551 & 0.542 & 0.620 & 25.7 & 2680 &670 &12.3
\\
MemOS  & 0.872 & 0.617 & 0.963 & 0.514 & 0.840 & 0.557 & 0.652 & 27.4 & 1682 & 206 & 2.3
\\
MemR3 & 0.929 & 0.929 & 0.600 & 0.684 & 0.769 & 0.639 & 0.742 & - & 6347 & 437 & 18.9 \\
GAM & 0.957 & 0.964 & 0.600 & 0.752 & 0.756 & 0.629 & 0.764 & - & 37850 & 1448 & 42.3 \\
\midrule
Ours (summary) & 0.900 & 0.857 & 0.833 & 0.549 & 0.756 & 0.519 & 0.678 & 19.6 & \textbf{380} & \textbf{11} & \textbf{5.37} \\
Ours (raw-data) & \textbf{0.971} & \textbf{0.982} & \textbf{0.900} & 0.752 & 0.846 & 0.662 & \textbf{0.808} & - & 7298 & 677 & 13.56 \\
\textbf{Ours (router)} & 0.957 & 0.946 & 0.833 & 0.692 & 0.744 & 0.602 & 0.752 & - & (4235+660) & (354+101) & 10.04 \\
Ours (router, 4.1-mini) & 0.929 & \textbf{0.982} & 0.867 & 0.714 & 0.782 & 0.624 & 0.770 & - & 4574 & 328 & 9.74 \\
\bottomrule
\end{tabular}
}
\label{tab:longmemeval_full_results}
\end{table*}

\paragraph{Why routing helps.}
A routed system is a mixture of two regimes: the cheap \textsc{Answer} path (Tier-1 only) and the expensive \textsc{Escalate} path (Tier-2 grounding).
The goal is therefore not to maximize escalation, but to:
(i) achieve high accuracy on the \textsc{Answer} path when summaries are semantically sufficient,
and (ii) reliably trigger \textsc{Escalate} when summaries are underspecified or high-risk. By dynamically detecting insufficiency, TierMem effectively converts the high UOR observed in baselines (15-30\%) into recoverable cases via provenance-guided escalation.
We quantify this in \S\ref{sec:analysis_ablations}.
\section{Analysis and Ablations}
\label{sec:analysis_ablations}

We analyze how TierMem achieves its accuracy--efficiency trade-off and isolate the contributions of its key mechanisms.
We focus on routing behavior (RQ2), provenance pointers (RQ3), and online consolidation (RQ4), and conclude with error analysis.

\subsection{Routing Behavior Breakdown}
\label{sec:routing_breakdown}
On LoCoMo, the final router (SFT+GRPO) escalates \textbf{39.0\%} of queries and recalls \textbf{71.7\%} of oracle-hard cases (Table~\ref{tab:router_learning}).
Routing overhead is \textbf{678 tokens/query} on average, decomposed as \textbf{584} input and \textbf{94.5} output tokens (Tok$_{\text{in}}{=}3396.4{+}584$, Tok$_{\text{out}}{=}247.7{+}94.5$; Table~\ref{tab:locomo_main}).
This corresponds to \textbf{14.7\%} of total input tokens (584/3980.4) and \textbf{15.7\%} of total tokens (678/4322.6) per query.
We analyze how training improves this routing boundary in \S\ref{sec:router_learning}.

\subsection{Router Learning and Cost--Fidelity Trade-off}
\label{sec:router_learning}
To isolate the impact of router training (RQ2), we compare several router variants in Table~\ref{tab:router_learning}.
We report both \emph{mechanism metrics} (routing classification) and \emph{outcome metrics} (end-to-end QA).
\paragraph{Training stages improve both behavior and overhead.}
The \textbf{Zero-shot} router is not only less accurate, but also inefficient due to poor protocol adherence, producing excessive output tokens.
\textbf{SFT} regularizes the output format and establishes basic routing logic, improving recall on hard queries while reducing overhead.
\textbf{SFT+GRPO} further sharpens the decision boundary: it improves hard-case recall to \textbf{71.7\%} and achieves the best local routing F1-score, indicating that RL improves discrimination rather than merely escalating more.

\paragraph{End-to-end impact.}
Despite a small routing overhead (Table~\ref{tab:router_learning}), \textbf{SFT+GRPO} matches the stronger GPT-4.1-mini router in overall LoCoMo accuracy (0.851) and reduces latency substantially compared to raw-only.
These results support the core premise of TierMem: a lightweight router can act as a semantic cache controller, allocating expensive raw evidence only when needed.

\begin{table*}[t]
\centering
\small
\caption{\textbf{Impact of Training Strategies on Router Behavior.}
\textbf{Router Cost} denotes the average token overhead (Input+Output) incurred by the routing step itself.
The \textbf{Zero-shot} baseline is inefficient, generating excessive output tokens ($\sim$317 avg output) due to poor instruction following.
\textbf{SFT} regularizes the output format, reducing overhead to $\sim$678 tokens.
\textbf{SFT+GRPO} achieves the best performance trade-off: for a negligible router overhead ($\sim$678 tokens), it recovers \textbf{71.7\%} of hard queries and matches GPT-4.1-mini's accuracy, effectively saving 51\% of total system costs compared to the Raw-only upper bound. Additional analysis of S/R counterfactual outcomes and why similar end-to-end accuracy can arise despite different hard-case recall is provided in Appendix~\ref{app:router_counterfactual}.
}
\resizebox{\textwidth}{!}{
\begin{tabular}{l|ccc|cccc}
\toprule
& \multicolumn{3}{c|}{\textbf{Routing Classification (Mechanism)}} & \multicolumn{4}{c}{\textbf{End-to-End QA Performance (Outcome)}} \\
\textbf{System} &
\textbf{R-rate} &
\textbf{Recall} (Hard) &
\textbf{F1-Score(\%)} &
\textbf{Overall Acc.} $\uparrow$ &
\textbf{Router Cost} $\downarrow$ &
\textbf{Router Lat.(s)}$\downarrow$ &
\textbf{Lat.(s)} $\downarrow$ \\
\midrule
Ours (Summary-only) & 0 \% & - & - & 0.755 & - & -& \textbf{1.24} \\
Ours (Raw-only) & 100\% & 100.0\% & - & \textbf{0.873} & - & - & 17.18 \\
\midrule
Ours (Zero-shot) & 20.6\% & 44.7\% & 38.1 & 0.806 & 901 & 1.541 & 5.41 \\
Ours (SFT) & 38.2\% & 65.8\% & 37.5 & 0.831 & 680 & 0.494 & 6.28 \\
\textbf{Ours (SFT+GRPO)} & 39.0\% & \textbf{71.7\%} & \textbf{40.6} & \underline{0.851} & 678 & 0.551 & 6.76 \\
\midrule
GPT-4.1-mini & 35.1\% & 76.4\% & 46.5 & 0.851 & 685 & 0.757 & 6.40 \\
\bottomrule
\end{tabular}
}
\label{tab:router_learning}
\end{table*}
\subsection{Ablation: Effect of Provenance Pointers}
\label{sec:ablation_provenance}
A key design choice in TierMem is explicit Tier-1$\rightarrow$Tier-2 provenance linking, used to warm-start escalation on likely relevant raw pages (RQ3).
We compare:
(i) \textbf{Linked}, full TierMem with provenance pointers enabled, and
(ii) \textbf{No-Linked}, an otherwise identical system where escalation performs global retrieval (BM25) without prioritizing linked raw pages.
Results are shown in Table~\ref{tab:provenance_main}.

\paragraph{Overall accuracy improves via better escalations.}
Provenance pointers yield a consistent end-to-end gain: \textbf{Linked} achieves \textbf{85.1\%} accuracy compared to \textbf{83.6\%} for \textbf{No-Linked} (+1.5pp), with similar routing rates.
Crucially, improvements concentrate on escalated queries: \textbf{Acc.@R} rises from \textbf{77.5\%} to \textbf{81.7\%} (+4.2pp), while the \textsc{Answer} path remains unchanged.
This matches the intended causal role of provenance pointers: they do not change what summaries contain, but they improve how quickly and reliably the system finds decisive Tier-2 evidence when summaries are insufficient.A more detailed breakdown of escalation behaviors and retrieval dynamics is provided in Appendix~\ref{app:provenance}.

\begin{table}[t]
\centering
\small
\setlength{\tabcolsep}{4pt}

\caption{
\textbf{Effect of provenance pointers.}
Linked enables provenance-aware warm-start during escalation.
Improvements are concentrated on escalated queries (Acc.@R), while summary-path accuracy remains unchanged.
}
\begin{tabular}{lccccc}
\toprule
\textbf{Method} &
\textbf{Acc.} &
\textbf{R-rate} &
\textbf{Acc.@R} &
\textbf{Tok$_{\text{in}}$/Q} &
\textbf{Lat.(s)} 
\\
\midrule
Raw-only & 87.3 & 100\% & 87.3 & 7398 & 17.18 \\
\midrule
No-Linked & 83.6 & 38.0\% & 77.5 & 3832 & 6.33\\
Linked (ours) & \textbf{85.1} & 39.0\% & \textbf{81.7} & 3980 & 6.76 \\
\bottomrule
\end{tabular}

\label{tab:provenance_main}
\end{table}

\begin{table*}[t]
\centering
\caption{\textbf{System evolution under consolidation in locomo(n=1540).}
\textbf{S-Traffic}:  queries routed to Summary (Tier-1 only).
\textbf{S-Acc.}: accuracy on Summary-routed queries.
\textbf{S-Correct}:  correctly answered via Summary.
\textbf{Tok$_{\text{avg}}$}/\textbf{Lat.}: average tokens/latency per query.}
\label{tab:system_evolution}
\renewcommand{\arraystretch}{1.18} 
\setlength{\tabcolsep}{9pt}      
\resizebox{1.8\columnwidth}{!}{   
\begin{tabular}{l|c|ccc|cc|cc}
\toprule
\textbf{Epoch} &
\textbf{Overall Acc.} &
\textbf{S-Traffic} &
\textbf{S-Acc.} &
\textbf{S-Correct} $\uparrow$ &
\textbf{Tok$_{\text{avg}}$} &
\textbf{Lat.(s)} &
\textbf{ADD} &
\textbf{UPDATE} \\
\midrule
\multicolumn{9}{l}{\textit{No-Recall write-back (no recall; only ADD or SKIP)}} \\
\midrule
\textbf{E1}  & 0.851 & 932   & 87.9\% & 819                       & 3,958 & 5.14 & 720 & -- \\
\textbf{E2}  & 0.847 & 1,125 & 87.6\% & 985 \scriptsize{(+166)}  & 3,035 & 4.16 & 486 & -- \\
\textbf{E3}  & 0.845 & 1,245 & 87.0\% & 1,083 \scriptsize{(+98)} & 2,419 & 3.39 & 317 & -- \\
\midrule
\multicolumn{9}{l}{\textit{Retrieve-and-Edit write-back (recall related Tier-1, then ADD/UPDATE/SKIP)}} \\
\midrule
\textbf{E1U} & 0.851 & 932   & 87.9\% & 819                       & 3,958 & 5.14 & 1,186 & 209 \\
\textbf{E2U} & 0.852 & 1,168 & 87.0\% & 1,016 \scriptsize{(+197)}& 2,719 & 4.22 & 652   & 105 \\
\textbf{E3U} & 0.850 & 1,224 & 86.8\% & 1,063 \scriptsize{(+57)} & 2,447 & 3.85 & 426   & 77 \\
\bottomrule
\end{tabular}
}
\end{table*}

\subsection{Consolidation Over Replay Epochs}
\label{sec:online_evolution}

TierMem supports continual improvement via \textbf{consolidation} (verified write-back).
For consistency with the benchmark protocol, we evaluate consolidation in an epoch-wise replay setting:
within each epoch, Tier-1 remains \textbf{fixed} while answering the full query set, and verified findings collected from escalations are written back \textbf{between} epochs.
This isolates the effect of consolidation on future queries while keeping per-epoch evaluation conditions identical (RQ4).

After a cache miss, TierMem escalates to Tier-2 to gather auditable evidence, then distills only \emph{verified} findings into Tier-1 as compact units \emph{linked} back to supporting raw pages.
Unlike generic summarization, write-back is \emph{query-triggered} and \emph{evidence-backed}: the system only commits what was needed and validated during escalation, revealing which details must persist under the write-before-query asymmetry.
In our implementation, the write-back controller supports three operations: \textbf{ADD}, \textbf{UPDATE}, and \textbf{SKIP}. 

\paragraph{Replay protocol (epoch-wise).}
We conduct a longitudinal study over three \textbf{replay epochs} using the same fixed query set.
Within each epoch, Tier-1 is \textbf{frozen}: answers are produced without modifying memory.
During the epoch, we \emph{log} the verified findings extracted from escalations.
After the epoch finishes, we apply these logged updates in a \textbf{single batch} to produce the next Tier-1 state for the following replay epoch.
We evaluate (i) overall task accuracy, (ii) the fraction of queries routed to the cheap Summary path (Tier-1 only), and (iii) efficiency (average input tokens and latency).
If consolidation is effective, later replay epochs should answer more queries from Tier-1 \emph{without losing correctness}, reducing the need for expensive raw grounding.

\paragraph{Two write-back variants.}
We compare two batch write-back strategies that differ in whether they \emph{recall} existing related Tier-1 units before committing a new fact:

\begin{itemize}
    \item \textbf{No-Recall write-back (E1--E3):} after completing an epoch, the system decides whether to write a new Tier-1 unit based only on each verified finding and its triggering query, without retrieving related Tier-1 units. If it writes, it always performs \textbf{ADD} (otherwise \textbf{SKIP}).
    \item \textbf{Retrieve-and-Edit write-back (E1U--E3U):} after completing an epoch, the system first retrieves semantically related Tier-1 units using each verified finding as a query, then chooses among \textbf{ADD/UPDATE/SKIP}. This enables \textbf{UPDATE} and avoids uncontrolled growth of redundant entries.
\end{itemize}

\paragraph{Results: consolidation increases cheap-path coverage and amortizes raw cost across replays.}
Table~\ref{tab:system_evolution} summarizes the evolution.
Across replay epochs, overall accuracy remains stable, yet the system answers more queries on the cheap summary path:
\textbf{S-Traffic} increases monotonically, and \textbf{S-Correct} rises by \textbf{32\%} (819$\to$1{,}083) under No-Recall write-back.
This indicates that as Tier-1 accumulates verified units \emph{from prior replays}, a growing fraction of queries become answerable with sufficient evidence from summaries alone, offloading work that previously required raw-page reads.
Correspondingly, average token usage and latency decrease over time (e.g., Tok$_{\text{avg}}$: 3{,}958$\to$2{,}419; Latency: 5.14s$\to$3.39s), demonstrating clear amortization of raw-access cost under repeated query demand.

\paragraph{Retrieve-and-Edit improves consolidation hygiene.}
Retrieve-and-Edit exhibits a similar trend of increasing Summary-path coverage, while maintaining overall accuracy at least as well as No-Recall at later epochs (e.g., 0.852 at E2U vs.\ 0.847 at E2).
Because it explicitly recalls related Tier-1 units, it can \textbf{UPDATE} rather than blindly append, reducing semantic duplication and keeping Tier-1 more coherent.
This comes with a modest management overhead: Tok$_{\text{avg}}$ decreases substantially, while latency can be slightly higher than No-Recall at the same epoch due to the additional recall-and-edit step.
Overall, Retrieve-and-Edit provides better consolidation hygiene while still enabling strong amortization.


\begin{table*}[t!]
\centering
\footnotesize
\renewcommand{\arraystretch}{1}
\setlength{\tabcolsep}{3pt}
\caption{Qualitative case studies illustrating how TierMem aligns retrieval granularity with query requirements.}
\label{tab:case_study_table}

\resizebox{0.95\textwidth}{!}{%
\begin{tabular}{p{0.15\textwidth} | p{0.35\textwidth} | p{0.45\textwidth}}
\toprule
\textbf{User Query} & \textbf{Tier-1 Summary (Compressed)} & \textbf{Router Decision \& Reasoning} \\
\midrule

\multicolumn{3}{c}{\textit{\textbf{Case 1: Mitigating Entity Loss (The ``Hero'' Case)}}} \\
\midrule
\textbf{Q:} ``What book recommendations has Joanna given to Nate?'' &
\textit{Context:} ``...Joanna recommended Nate to find a \textbf{fantasy book series}... Joanna approved the \textbf{specific book series} Nate showed her...'' \newline
\textcolor{red}{\xmark \textbf{Defect:}} The specific entity (Book Title) is lost during summarization. &
\textbf{Decision:} \colorbox{orange!30}{\textsc{Escalate}} (Raw Search) \newline
\textbf{Router Thought:} ``Summaries mention Joanna recommended a series... but \textbf{do not explicitly state the book titles}. They are contextually related but lack specific answers.'' \newline
\textbf{Outcome:} Retrieved raw log containing ``\textit{Little Women}'', avoiding a vague or hallucinated answer. \\
\midrule

\multicolumn{3}{c}{\textit{\textbf{Case 2: Semantic Sufficiency (The ``Saver'' Case)}}} \\
\midrule
\textbf{Q:} ``What important values does John want to teach his kids through adopting a rescue dog?'' &
\textit{Context:} ``...John said adopting a rescue dog... teaches John's kids \textbf{responsibility and compassion}...'' \newline
\textcolor{green!60!black}{\cmark \textbf{Sufficient:}} Key semantic concepts are present. &
\textbf{Decision:} \colorbox{green!30}{\textsc{Answer}} (Summary Only) \newline
\textbf{Router Thought:} ``Summaries do not explicitly state `values: X', but they hint at key points... \textbf{Since summaries contain the answer} (responsibility and compassion), use Summary.'' \newline
\textbf{Outcome:} Answered correctly (Accuracy preserved) while skipping $\sim$4k tokens of raw retrieval. \\
\bottomrule
\end{tabular}%
}
\end{table*}

\subsection{Qualitative Case Studies}
\label{sec:case_study}
\subsubsection{Case Study: Qualitative Cases}
We present two representative cases illustrating how TierMem aligns retrieval granularity with query requirements in Table~\ref{tab:case_study_table}.
Case 1 requires escalation due to entity loss in summarization, and Case 2 can be answered cheaply from summaries because key semantic evidence is preserved. A broader clustering-based analysis of router decision errors is provided in Section~\ref{sec:router_error_analysis}.

\paragraph{Mitigating compression loss.}
When the user query requires a specific entity that is missing from Tier-1 summaries, TierMem escalates and grounds the answer in Tier-2 raw logs, avoiding vague or hallucinated responses.

\paragraph{Ensuring semantic efficiency.}
When Tier-1 summaries contain semantically sufficient evidence (even if not phrased in the same surface form as the query), the router answers directly, avoiding unnecessary raw retrieval while preserving correctness.

\subsubsection{Anatomy of Summary-Only Failures and Router Decisions}
\label{sec:summary_failures_router}

To better understand the \emph{write-before-query} barrier and how our router responds to summary insufficiency, we analyze the subset of test queries where the \textbf{Summary-Only} system fails.
On LoCoMo ($n{=}1540$), Summary-Only makes \textbf{378} errors (\textbf{24.5\%}).
These errors expose two structural failure modes of write-time compression:
(i) \textbf{information is missing} from Tier-1 (no recoverable answer span), and
(ii) \textbf{over-generalization} where decisive entities/values are replaced by coarse abstractions (e.g., ``Sweden'' $\rightarrow$ ``her home country'').
Table~\ref{tab:summary_only_root_causes} summarizes the dominant root causes.

\paragraph{Write-before-query manifests as missing or unrecoverable evidence.}
The largest portion of Summary-Only errors (\textbf{41.3\%}) are due to \textbf{missing information}:
the needed answer is not present in the retrieved summaries at inference time.
This is a direct symptom of write-before-query: the memory writer must decide what to keep before knowing the future query, so some decisive constraints/attributes are never preserved in Tier-1 in a query-usable form.
A second major slice (\textbf{28.8\%}) is \textbf{over-generalization}, where summaries retain topic relevance but lose \emph{answer-critical specificity} (entities, modifiers, exact values), making the evidence \emph{semantically related yet insufficient}.

\paragraph{Router decisions on Summary-Only failures.}
Given these 378 Summary-Only errors, our router routes \textbf{72.0\%} (272) to \textsc{Escalate} (R) and \textbf{28.0\%} (106) to \textsc{Answer} (S).
However, only \textbf{10/106} (\textbf{9.4\%}) of S decisions are actually justified by summaries containing the gold information;
the remaining \textbf{96/106} are false cache hits where the router should have escalated.

\paragraph{Why many errors persist even after escalation.}
A key finding is that \textbf{72.0\%} of Summary-Only failures are \emph{already} recognized as summary-insufficient (router chooses R), yet the final answer remains wrong.
This indicates that remaining failures are dominated by the \textbf{R-path} rather than the routing decision itself:
(i) retrieval may fail to locate the correct raw page,
(ii) the raw evidence may still be missing or already generalized upstream (i.e., the source-of-truth does not contain an explicit answer span in the stored pages), or
(iii) the generator may fail to extract/compose the correct answer even when relevant evidence is present.
This motivates improving \emph{provenance-guided retrieval quality} and \emph{grounded answer extraction} in addition to improving the router.

\begin{table*}[t]
\centering
\small
\setlength{\tabcolsep}{10pt}
\caption{\textbf{Root causes of Summary-Only errors} on LoCoMo ($n{=}1540$; Summary-Only errors $=378$).
The top two categories (missing information and over-generalization) are direct manifestations of the write-before-query barrier.}
\label{tab:summary_only_root_causes}
\begin{tabular}{lrrp{1\columnwidth}}
\toprule
\textbf{Root cause} & \textbf{\#} & \textbf{\%} & \textbf{Description} \\
\midrule
Information missing & 156 & 41.3 & Retrieved summaries contain no explicit answer evidence (compression/coverage loss). \\
Over-generalization & 109 & 28.8 & Specific answer-critical details are abstracted away (e.g., entity/value $\rightarrow$ generic descriptor). \\
Reasoning failure & 43 & 11.4 & Requires multi-step reasoning but summaries do not support it. \\
Routing decision error & 20 & 5.3 & Router selects S when summaries are insufficient (false cache hit). \\
Time confusion & 11 & 2.9 & Multiple similar events/timestamps induce ambiguity. \\
Entity confusion & 6 & 1.6 & Entity references collapse to pronouns or ambiguous mentions. \\
Other & 33 & 8.7 & Miscellaneous failures. \\
\bottomrule
\end{tabular}
\end{table*}

\subsection{Router Diagnostics on Summary-Only Error Cases}
\label{subsec:router_diagnostics_on_error_cases}

We further analyze the router system on the same 431 Summary-Only error cases to diagnose whether failures stem from (i) incorrect \emph{routing decisions} or (ii) limitations in the \textsc{Escalate} (raw-grounded) path.

\paragraph{A large fraction are false cache hits (should escalate but answered from summaries).}
Among the 431 Summary-Only errors, the router selects \textsc{Answer} in \textbf{191} cases (44.3\%).
Crucially, \textbf{148} of these are \textbf{false cache hits}---the router answers from Tier-1 although summaries do \emph{not} contain the gold evidence (\textbf{34.3\%} of all 431 error cases).
This indicates that the dominant router failure mode is \emph{insufficient strictness} in sufficiency checking, often triggered by superficial keyword matches that do not verify slot completeness or exactness.

\paragraph{Escalation is often triggered correctly but can still fail downstream.}
In \textbf{240} cases (55.7\%), the router correctly escalates, yet the final answer remains wrong.
This suggests that improvements must also target the \textsc{Escalate} pipeline, e.g., ensuring the raw store contains the needed detail, improving provenance-guided retrieval coverage, and strengthening the grounding constraints during generation to avoid falling back to the same generalized phrasing.

\paragraph{Connecting back to write-before-query.}
These results highlight a two-part mitigation strategy for write-before-query:
(1) \textbf{reduce lossy abstraction at write time} (e.g., entity/number/date retention constraints), and
(2) \textbf{tighten inference-time sufficiency checks} to detect when summaries are merely topically related but lack decisive evidence.
In our setting, (2) is especially urgent, as false cache hits directly harm correctness.

\begin{table}[t]
\centering
\small
\setlength{\tabcolsep}{6pt}
\caption{\textbf{Router behavior on Summary-Only error cases (n=431).}
False cache hits (choosing \textsc{Answer} when escalation is needed) account for 34.3\% of the error set.
}
\label{tab:router_on_summary_only_errors}
\begin{tabular}{lrr}
\toprule
\textbf{Router outcome on the error set} & \textbf{Count} & \textbf{Rate} \\
\midrule
Escalate but still wrong  & 240 & 55.7\% \\
False cache hit  & 148 & 34.3\% \\
Answer with summaries and sufficient  & 43 & 10.0\% \\
\bottomrule
\end{tabular}
\end{table}

\section{Related Work}

\subsection{Memory for Long-Horizon LLM Agents}
Long-horizon LLM agents must remain consistent over histories spanning days to months, yet context windows and inference budgets make feeding full raw logs costly and sometimes ineffective: even with context-extension methods, compute grows with length, and models exhibit position-sensitive degradation on long inputs \cite{chen_extending_2023,liu_lost_2023}.
Accordingly, non-parametric agent memory systems rely on external read--write stores to support long-term personalization \cite{yehudai_survey_2025}.
Representative approaches include OS-style hierarchical management (MemGPT) and controller-based memory operation (SCM), as well as write-time compression and organization (MemoryBank, Mem0, LightMem) and structured memories such as temporal knowledge graphs \cite{packer_memgpt_2024,wang_scm_2023,zhong_memorybank_2023,chhikara_mem0_2025,fang_lightmem_2025,rasmussen_zep_2025}.
In contrast, TierMem explicitly treats memory access as inference-time evidence granularity allocation: it answers from a cheap, provenance-linked summary tier when sufficient, and escalates to an immutable raw tier only on detected insufficiency, with verified write-back for amortization.
We evaluate this trade-off on long-horizon benchmarks including LoCoMo and LongMemEval \cite{maharana_evaluating_2024,wu_longmemeval_2025}.

\subsection{Agentic Memory Retrieval and Auditable Grounding}
Beyond storage, retrieval in long-term interaction is inherently a control problem: models must decide when and how much to retrieve and when to stop to avoid under/over-retrieval \cite{asai_self-rag_2023,jiang_active_2023}.
Related work studies iterative retrieval and reflection-based improvement, \cite{shinn_reflexion_2023} and emphasizes verifiable grounding via retrieve--revise pipelines and fine-grained citations \cite{gao_rarr_2022,cao_verifiable_2024}.
Closer to agent memory, retrieval controllers such as MemR$^3$ and page-store-based deep research systems optimize closed-loop evidence gathering under cost constraints \cite{du_memr3_2025,yan_general_2025}.
In contrast, TierMem collapses control into a cost-sensitive binary routing decision (\textsc{Answer} vs.\ \textsc{Escalate}) across memory granularities, uses provenance pointers to enable auditable fallback to raw pages, and consolidates only verified evidence back to the fast tier.

\section{Conclusion}
\label{sec:conclusion}

We identify a structural failure mode in summary-centric long-horizon memory: the \textbf{write-before-query} barrier, where write-time compression can induce \textbf{unverifiable omissions} and break auditability.
To address this, we propose \textbf{TierMem}, which treats retrieval as \textbf{inference-time evidence allocation}: answer from a fast summary tier when evidence is sufficient, and \textsc{Escalate} to an immutable raw-log source of truth only when necessary, then write back \emph{verified} findings with provenance links.

Across \textbf{LoCoMo} and \textbf{LongMemEval}, TierMem improves the accuracy--efficiency tradeo-off, approaching raw-grounded faithfulness while substantially reducing query-time cost.
Analyses further show that (i) a lightweight sufficiency router can reliably control escalation with small overhead, (ii) provenance pointers materially improve escalated grounding quality, and (iii) verified write-back increases cheap-path coverage over time without degrading accuracy.

\bibliography{references}
\bibliographystyle{icml2026}

\appendix
\section{Implementation Details}
\label{app:impl_details}

\subsection{Memory Instantiation}
Our framework is general; in experiments we instantiate the memory management system using the open-source \textbf{Mem0} library~\cite{chhikara_mem0_2025} for vector indexing and storage. We adapt Mem0 to our \textbf{Paged-Log Architecture} as follows:
\begin{itemize}
    \item \textbf{Summarization prompt ($f_{\text{summ}}$).}
    We replace Mem0's default entity-graph extraction with \textbf{page-level summarization}. The prompt instructs the model to generate a concise, self-contained summary of the interaction log, preserving key dates, decisions, and user preferences.

    \item \textbf{Storage schema.}
    Each page summary is embedded and indexed in the vector store. We extend Mem0 metadata to include the Page ID ($P_i$) as a provenance pointer $\rho_i$, enabling retrieval to map back from a summary vector to its originating page.

    \item \textbf{Embedding model.}
    We embed summaries with \texttt{text-embedding-3-small}.
\end{itemize}

\paragraph{Mem0 components used vs. disabled.}
In our implementation, Mem0 is used only for (i) storing and querying embeddings for Tier-1 summary entries and (ii) attaching/retrieving metadata (e.g., \texttt{raw\_log\_id/page\_id}) for provenance.
We do not use Mem0's entity-graph construction, graph updates, or graph-based retrieval; these components are disabled/bypassed so that Tier-1 consists solely of page summaries plus metadata pointers.

\subsection{Paged-Log Ingestion and Persistence}
\label{app:paged_log_ingestion}
We implement a \textbf{paged ingestion} procedure that converts streaming dialogue turns into fixed-size \emph{pages} stored in a persistent \texttt{PageStore}. Each incoming turn is appended to the current page with an explicit timestamp and speaker tag. Once a page reaches a configured maximum length (default: 1000 tokens), we trigger a summarization-and-indexing step via \texttt{mem0.add(infer=True)} over the entire page.

\paragraph{Page format.}
Each page stores raw interaction text line-by-line in the canonical form:
\[
[\text{timestamp}]~\text{speaker}:~\text{text}.
\]
This preserves multi-speaker context for datasets with more than one interlocutor.

\paragraph{Provenance.}
For each page $P_i$, we generate a unique \texttt{page\_id}. We set \texttt{raw\_log\_id} $\leftarrow$ \texttt{page\_id} in Mem0 metadata such that any retrieved summary hit can be deterministically mapped back to the original raw page content in \texttt{PageStore}.

\paragraph{Durability and replay.}
\texttt{PageStore} persists pages to disk and supports session-scoped loading. This enables resuming runs without re-ingesting all prior turns and ensures that the raw evidence used for research-mode grounding remains accessible.

\subsection{Deferred Summarization for Unindexed Pages}
\label{app:auto_summary}
To handle sessions that terminate before a page fills (or failures during ingestion), we provide a \textbf{deferred summarization} routine \texttt{auto\_summary}. It enumerates pages whose summaries have not yet been stored and calls \texttt{mem0.add(infer=True)} for each such page. The returned memory strings are recorded in \texttt{PageStore} and the page is marked as indexed, ensuring that the vector store remains consistent with the persistent log.

\subsection{Query Answering Pipeline}
\label{app:qa_pipeline}
Given a query $q$, we first retrieve top-$k$ summary hits from Mem0 using vector similarity search. Each hit contains a summary text and a provenance pointer to its source page.

\paragraph{Two-path routing (S/R).}
We use a lightweight router to select between:
\begin{itemize}
    \item \textbf{S-path (Summary-only).} When retrieved summaries contain an explicit answer, we generate the response directly from the summaries.
    \item \textbf{R-path (Research).} When summaries are ambiguous or incomplete, we switch to a slower research mode that grounds generation in raw page evidence.
\end{itemize}

\paragraph{Cost accounting.}
We record (i) Mem0 retrieval token usage and latency, (ii) router token usage, and (iii) answer-generation token usage. These are aggregated into online cost metrics for analysis.

\subsection{Research Mode: Integration--Plan Loop }
\label{app:research_v2}
The R-path implements an iterative loop that alternates between \textbf{integration} (extracting grounded facts from evidence) and \textbf{planning} (deciding whether to retrieve more evidence).

\paragraph{Evidence construction.}
We group retrieved hits by \texttt{page\_id} and build an evidence block that includes:
(i) Mem0 page summaries (when available) and
(ii) the full raw page content from \texttt{PageStore}.
This ensures that research mode always has access to verbatim text for grounding.

\paragraph{Integration step.}
The integration prompt extracts a set of \emph{linked facts} of the form:
\[
\texttt{fact},~\texttt{evidence\_quote}.
\]
Crucially, we \emph{do not} ask the model to generate provenance identifiers.

\paragraph{Deterministic provenance linking.}
After integration, we map each extracted fact to its source pages using a deterministic post-processing step:
\begin{itemize}
    \item First, we attempt substring matching of the normalized \texttt{evidence\_quote} against each page's raw content (and, if needed, its summaries).
    \item If quote matching fails, we fall back to keyword overlap between the fact text and page content.
\end{itemize}
This produces \texttt{source\_pages} and \texttt{evidence\_snippets} without relying on model-generated page identifiers, reducing hallucinated provenance.

\paragraph{Planning step.}
A separate planning prompt evaluates coverage and returns either \texttt{DONE} or \texttt{SEARCH} with retrieval commands. We support two retrieval operators:
\begin{itemize}
    \item \textbf{MEM0\_SEARCH}: semantic retrieval over Mem0 summaries.
    \item \textbf{KEYWORD\_SEARCH}: BM25-style keyword search over raw pages in \texttt{PageStore}.
\end{itemize}
We deduplicate issued commands and terminate early when no new pages are discovered.

\subsection{Dynamic Write-back Protocol}
\label{app:writeback}

We implement online consolidation as a closed-loop memory update step. Given a set of \emph{verified} linked facts extracted during the Research Integration phase:

\paragraph{1. Conflict Retrieval.}
For each new candidate fact $f_{new}$, we query $\Msum$ to retrieve the top-$k$ (e.g., $k=3$) most similar existing summary units, denoted as $\mathcal{M}_{retrieved}$.

\paragraph{2. Operation Selection.}
We prompt an LLM (the \textit{Memory Manager}) to compare $f_{new}$ with $\mathcal{M}_{retrieved}$. The model outputs one of four actions:
\begin{itemize}
    \item \texttt{SKIP}: If $f_{new}$ is semantically entailed by $\mathcal{M}_{retrieved}$.
    \item \texttt{UPDATE}: If $f_{new}$ adds specific details (e.g., timestamps, names) to a vague existing entry. The model generates the merged text.
    \item \texttt{ADD}: If $f_{new}$ represents a distinct, previously unindexed event.
\end{itemize}

\paragraph{3. Provenance Inheritance.}
If an \texttt{UPDATE} occurs, the new unit inherits the provenance pointers $\rho$ from both the original unit and the new evidence sources, maintaining a complete audit trail.
\subsection{Reranking (Optional)}
\label{app:reranking}
When enabled, we apply a \textbf{page-level reranker} to candidate evidence pages. Hits are first grouped by \texttt{page\_id}; reranking operates on each page's raw content (or, if missing, concatenated summaries) and selects the top-$k$ \emph{unique pages}. We also include a \textbf{protection rule}: pages with high-confidence Mem0 similarity scores are retained to avoid discarding highly relevant memories before reranking.

\paragraph{Discussion: Online Consolidation vs. Epoch-wise Replay Evaluation}

TierMem is designed to support \emph{online} consolidation: after an escalation, verified findings can be written back immediately to reduce future misses.
However, to match the benchmark evaluation protocol and avoid within-run distribution shift, our main experiments report results under an epoch-wise replay setting where Tier-1 is held fixed within each epoch and updates are applied between epochs.
We expect that in interactive deployments, moving write-back into the QA loop is straightforward and may further reduce repeated escalations, but careful design is needed to preserve reproducibility and to avoid unintended interactions with evaluation sampling.

\subsection{Statistics and Debug Instrumentation}
\label{app:stats}
For analysis, we log per-query events including (i) route choice (S vs.\ R), (ii) hit breakdown by \texttt{source\_type} (\texttt{original} vs.\ \texttt{linked\_fact}), (iii) linked-fact write-back counts, and (iv) latency/token usage. This instrumentation is used only for evaluation and does not affect the core algorithm.

\begin{algorithm}[t]
\caption{TierMem QA (Abstract): Router + Bounded Retrieve--Integrate--Plan Loop}
\label{alg:TierMem_qa_retrieve_integrate_plan}
\begin{algorithmic}[1] 
\REQUIRE Query $Q$; Tier-1 summaries $\Msum$; Tier-2 raw pages $\Mraw$
\REQUIRE Router $\pi_\theta$; generator $\mathrm{Gen}$
\REQUIRE $\mathrm{Retrieve}, \mathrm{Rerank}, \mathrm{Integrate}, \mathrm{Plan}$
\REQUIRE max iterations $T_{\max}$
\ENSURE Answer $\hat{Y}$

\STATE $Z_0 \gets \mathrm{Retrieve}(\Msum, Q)$
\IF{$\pi_\theta(Q, Z_0)=\textsc{Answer}$}
    \STATE \textbf{return} $\mathrm{Gen}(Q, Z_0)$
\ENDIF

\STATE $E \gets \mathrm{Integrate}(Q, \mathrm{Linked}(Z_0))$ \COMMENT{seed: summaries + linked raw pages (compressed)}
\STATE $q_1 \gets Q$; $H \gets \emptyset$ \COMMENT{optional: searched queries / history}

\FOR{$t=1$ \textbf{to} $T_{\max}$}
    \STATE $\mathcal{C}_t \gets \mathrm{Retrieve}(\Msum, q_t)\ \cup\ \mathrm{Retrieve}(\Mraw, q_t)$
    \STATE $X_t \gets \mathrm{TopK}\big(\mathrm{Rerank}(\mathcal{C}_t, Q)\big)$
    \STATE $E \gets E \cup \mathrm{Integrate}(Q, X_t)$ 
    \STATE $(d_t, q_{t+1}) \gets \mathrm{Plan}(Q, E, H)$ \COMMENT{$d_t\in\{\textsc{Stop},\textsc{Search}\}$}
    \IF{$d_t=\textsc{Stop}$} \STATE \textbf{break} \ENDIF
    \STATE $H \gets H \cup \{q_t\}$; $q_t \gets q_{t+1}$
\ENDFOR

\STATE $\hat{Y} \gets \mathrm{Gen}(Q, E)$
\STATE \textbf{optional:} $\mathrm{WriteBack}(\Msum, E)$
\STATE \textbf{return} $\hat{Y}$
\end{algorithmic}
\end{algorithm}




\section{Router: Data Construction, Training Pipeline, and Hyperparameters}
\label{app:router_full}

This appendix consolidates all details related to router supervision, SFT distillation, GRPO optimization, and the exact hyperparameters used.

\subsection{Task Definition and Notation}
\label{app:router_task}

The router is a lightweight controller that decides whether the system can answer a user query using Tier-1 summaries alone (\textsc{Answer}/S) or must escalate to Tier-2 raw logs (\textsc{Escalate}/R).
Given a query $Q$ and an initial evidence state $Z_0$ consisting of the top-$k$ retrieved Tier-1 summaries, the router predicts:
\[
a \sim \pi_\theta(a \mid Q, Z_0), \qquad a \in \{\texttt{S},\texttt{R}\}.
\]
The router output is \emph{strictly} a JSON object with an action token (and optionally a teacher \texttt{thinking} field during distillation).

\subsection{Data Construction: Summary Sufficiency Labels}
\label{app:router_data}

\paragraph{Source benchmark slice.}
We construct router training data from the \emph{Long-Range Understanding (LRU)} competency in MemoryAgentBench~\cite{hu_evaluating_2025}, using a summarization-style subset of 2,000 instances. Each instance is materialized into a linked two-tier memory state $(\Msum,\Mraw)$ using the write path described in \S\ref{sec:arch}.
For each query $Q$, Tier-1 dense retrieval returns the initial summary evidence $Z_0$ (top-$k_s$ summaries), which forms the router input.

\paragraph{Two-path execution (Summary vs.\ Research).}
For each query $Q$, we execute two fixed policies to determine whether $Z_0$ is sufficient:
\begin{itemize}
  \item \textbf{Summary-only (S-path):} $\hat{Y}_S \leftarrow \mathrm{Gen}(Q, Z_0)$
  \item \textbf{Research/raw-grounded (R-path):} Run the bounded Deep Search Loop (\S\ref{sec:protocol}) to obtain $E_{\mathrm{final}}$, then $\hat{Y}_R \leftarrow \mathrm{Gen}(Q, E_{\mathrm{final}})$
\end{itemize}

\paragraph{LLM judge and sufficiency rubric.}
\label{app:router_judge}
We obtain binary indicators $c_S, c_R \in \{0,1\}$ using LLM-based judges with conservative rubrics.
Both judges are run with temperature $0$ and are constrained to output a fixed JSON schema to reduce variance.

\paragraph{S-path: summary sufficiency judge.}
Given $(Q, y^\star, Z_0)$, the judge determines whether the retrieved summaries contain \emph{explicitly sufficient} information to answer the question correctly \emph{without guessing}.
We use the following prompt (verbatim):

\subsection{Judge Prompt for Summary Sufficiency}
\label{app:llm_as_judge}

\begin{figure}[htb]
\begin{AIbox}[width=\columnwidth]{Judge Prompt (Summary Sufficiency)}
\footnotesize
You are evaluating whether retrieved summaries contain sufficient information to answer a question.

Question: \{question\} \\
Gold answer: \{gold\_answer\}

Retrieved summaries: \\
\{summaries\_text\}

\textbf{Your task:} Determine if the summaries contain EXPLICIT information to answer the question correctly.

\textbf{Strict criteria (be conservative):}
\begin{enumerate}
    \item The answer should be DIRECTLY stated or clearly derivable from the summaries.
    \item Do NOT count vague/related information as sufficient.
    \item Do NOT infer causes from effects.
    \item For completeness questions (``how many'', ``list all'', ``both''), summaries must be COMPLETE.
    \item For exact details (dates, numbers, names), summaries must contain those exact details.
\end{enumerate}

\textit{If you have ANY doubt whether the summaries allow a factual answer without guessing, answer false.}

Output JSON:
\begin{verbatim}
{
  "has_sufficient_info": true/false, 
  "reason": "..."
}
\end{verbatim}
\end{AIbox}
\caption{The judge prompt used to evaluate whether the retrieved summaries provide sufficient context to answer the ground truth.}
\label{fig:judge_sufficiency_prompt}
\end{figure}

We set $c_S=\mathbb{I}[\texttt{has\_sufficient\_info}=\texttt{true}]$.

\paragraph{R-path: research/correctness judge.}
Given $(Q, y^\star, \hat{Y}_R)$, the judge determines whether the research-path answer matches the gold answer.
We use a separate LLM-as-judge prompt (Appendix~\ref{app:llm_as_judge}), which outputs a JSON object of the form:
\{\,"label": \texttt{``CORRECT''} / \texttt{``WRON''}\,\}.
We set $c_R=\mathbb{I}[\texttt{label}=\texttt{``CORRECT''}]$.

\paragraph{Why conservative judging?}
Our routing label targets \emph{summary sufficiency}.
False positives (marking summaries as sufficient when they are not) would train a router that under-escalates and produces ungrounded answers.
Therefore, the S-path prompt intentionally biases toward \texttt{false} when there is ambiguity or missing explicit support.

\paragraph{Routing labels (semantic cache hit/miss).}
We define the router supervision label as:
\[
y =
\begin{cases}
\texttt{S} & \text{if } c_S = 1,\\
\texttt{R} & \text{if } c_S = 0 \land c_R = 1.
\end{cases}
\]
We drop instances where both policies fail ($c_S{=}0 \land c_R{=}0$), since they provide no signal on whether escalation can recover missing evidence.

\subsection{Training Workflow Overview}
\label{app:router_workflow}

We train the router using a two-stage pipeline:
\begin{enumerate}
  \item \textbf{Stage 1 (SFT / Distillation):} Initialize the router to follow the routing protocol and learn strong sufficiency heuristics from a high-capacity teacher (GPT-5) using a strict JSON format.
  \item \textbf{Stage 2 (GRPO):} Starting from the SFT checkpoint, refine the decision boundary under an explicit accuracy--cost trade-off, discouraging unnecessary escalation while preserving faithfulness.
\end{enumerate}
This separation is intentional: SFT primarily regularizes behavior and format; GRPO optimizes the accuracy--efficiency trade-off.

\subsection{Stage 1: SFT via GPT-5 Thinking+Action Distillation}
\label{app:router_sft}

\paragraph{Teacher prompt (GPT-5 thinking trace + binary action).}
We distill a teacher-generated \texttt{thinking} trace and the final routing decision \texttt{action} using the following prompt (verbatim):

\subsection{Router Thinking Prompt}
\label{app:router_prompt_v2}

\begin{figure*}[htb]
\begin{AIbox}[width=\textwidth]{Router Thinking Prompt }
\footnotesize
You are a router for a memory-augmented QA system. Decide whether the retrieved summaries are sufficient to answer the question.

\textbf{Available Actions}
\begin{enumerate}
    \item \textbf{``S'' — Answer using current summaries only} \\
    Use when: The summaries contain a span that directly states the answer (or a direct paraphrase) WITHOUT needing to assume missing facts. \\
    \textit{Rules of thumb:}
    \begin{itemize}
        \item If you can point to a specific snippet that can be copied into the answer, choose S.
    \end{itemize}

    \item \textbf{``R'' — Deep research mode} \\
    Use when: Summaries are too ambiguous, contradictory, or unrelated; OR the question requires a comprehensive list and summaries are clearly incomplete. \\
    R is more reliable than S.
\end{enumerate}

\textbf{Input} \\
QUESTION: \{question\} \\
SUMMARIES: \\
\{summaries\_block\}

\textbf{Decision checklist}
\begin{itemize}
    \item What exact answer format is needed? (single value / list / date / identity)
    \item Do summaries explicitly contain it?
    \item Any ambiguity or need to infer?
    \item Any completeness requirement?
\end{itemize}

\textbf{Output (JSON only)}
\begin{verbatim}
{
  "thinking": "<your reasoning>",
  "action": "S/R"
}
\end{verbatim}
\end{AIbox}
\caption{The updated router prompt incorporating explicit chain-of-thought reasoning to improve decision reliability.}
\label{fig:router_prompt_v2}
\end{figure*}
\paragraph{SFT targets and formatting.}
For each example, the student router is trained to reproduce the teacher's JSON response.
We treat the entire JSON (including \texttt{thinking} and \texttt{action}) as the supervised target, using standard next-token cross-entropy.
At inference time, the system uses only the \texttt{action} field.

\paragraph{Teacher-label consistency filter.}
To ensure alignment with oracle labels derived in \S\ref{app:router_data}, we keep only SFT examples where the teacher action matches the oracle routing label $y$.
Examples with mismatched teacher actions are discarded.

\paragraph{Why include thinking in SFT.}
The teacher \texttt{thinking} field provides structured intermediate reasoning about evidence sufficiency (e.g., completeness, ambiguity, exactness), which improves stability and reduces instruction-following failures (e.g., verbose or malformed outputs) in a small router.

\subsection{Stage 2: GRPO Alignment for Accuracy--Cost Trade-off}
\label{app:router_grpo}

Starting from the SFT checkpoint, we optimize the router with GRPO using a trajectory-level reward:
\begin{equation}
R(\tau) = R_{\mathrm{acc}}(\tau)\;-\;\lambda_{\mathrm{cost}} R_{\mathrm{cost}}(\tau)\;-\;\lambda_{\mathrm{waste}} R_{\mathrm{waste}}(\tau),
\label{eq:app_router_reward}
\end{equation}
where:
\begin{itemize}
  \item $R_{\mathrm{acc}}$ encourages correct routing (choose S when summaries suffice; choose R when escalation is necessary),
  \item $R_{\mathrm{cost}}$ penalizes choosing R to reflect token/latency overhead,
  \item $R_{\mathrm{waste}}$ penalizes \emph{unnecessary escalation} (choosing R when $y=\texttt{S}$).
\end{itemize}

\paragraph{Concrete reward implementation.}
In our implementation, rewards are computed from $(c_S, c_R)$ and the router decision $a$:

\paragraph{GRPO dataset construction.}
We build GRPO prompts using the same router input $(Q, Z_0)$ as in SFT. For each prompt, we sample multiple router completions per prompt (generations per prompt is reported in \S\ref{app:router_hparams}) to compute group-relative advantages.

\subsection{GRPO Dataset Preparation and Filtering}
\label{app:router_grpo_data}

Starting from the offline QA logs (which include both summary-path and research-path outcomes), we apply the following filters:
\begin{enumerate}
  \item Remove instances where \textbf{both} paths fail ($c_S=0 \land c_R=0$).
  \item Remove instances where the \textbf{R-path fails} ($c_R=0$), since escalation does not recover evidence.
  \item Optionally re-judge S-path sufficiency using a fixed judge configuration to reduce label noise; keep only instances consistent under re-judging.
\end{enumerate}

\paragraph{Optional augmentation.}
We optionally augment a subset of questions by paraphrasing while preserving intent and specificity (2--3 paraphrases per question). Paraphrases inherit the same memory state and labels.

\paragraph{Balancing.}
To emphasize hard decisions, we oversample cases where escalation is necessary ($c_S=0 \land c_R=1$) during GRPO minibatch formation.

\subsection{Hyperparameters}
\label{app:router_hparams}

\subsubsection{SFT Hyperparameters}
\label{app:router_sft_hparams}

\begin{table}[H]
\centering
\small
\begin{tabular}{ll}
\toprule
Parameter & Value \\
\midrule
Train type & LoRA \\
Precision & bfloat16 \\
Epochs & 5 \\
Learning rate & $1\times 10^{-4}$ \\
Max sequence length & 4096 \\
Per-device train batch size & 8 \\
Per-device eval batch size & 4 \\
Gradient accumulation steps & 1 \\
LoRA rank / alpha & 64 / 128 \\
Target modules & all-linear \\
Warmup ratio & 0.1 \\
Eval / Save steps & 10 / 10 \\
Save total limit & 5 \\
Logging steps & 1 \\
Dataloader workers & 4 \\
Loss scaling & ignore\_empty\_think \\
Kernel optimization & use\_liger\_kernel=true \\
DeepSpeed & ZeRO-2 \\
\bottomrule
\end{tabular}
\caption{SFT (LoRA) training configuration for the router.}
\end{table}

\subsubsection{GRPO Hyperparameters}
\label{app:router_grpo_hparams}

\begin{table}[H]
\centering
\small
\begin{tabular}{ll}
\toprule
Parameter & Value \\
\midrule
RLHF type & GRPO \\
Initialization & SFT adapter checkpoint \\
Precision & bfloat16 \\
Epochs & 6 \\
Context length & 2048 \\
Learning rate & 5e-6 \\
Max completion length & 256 \\
Per-device train batch size & 24 \\
Per-device eval batch size & 24 \\
Gradient accumulation steps & 1 \\
Warmup ratio & 0.05 \\
DeepSpeed & ZeRO-2 \\
Target modules & all-linear \\
LoRA rank / alpha & 64 / 128 \\
Dataloader workers & 4 \\
Logging / Save steps & 5 / 50 \\
Completions logging & log\_completions=true \\
Sampling temperature & 1.2 \\
Generations per prompt & 8 \\
Rollout forward batch size & 8 \\
KL / regularization parameter & $\beta=0.02$ \\
Reporting & Weights \& Biases \\
\bottomrule
\end{tabular}
\caption{GRPO training configuration for router policy optimization.}
\end{table}

\subsubsection{Reward Coefficients}
\label{app:router_reward_coeffs}

\begin{table}[H]
\centering
\small
\begin{tabular}{ll}
\toprule
Name & Value \\
\midrule
Correct reward (\texttt{CORRECT\_REWARD}) & 1.0 \\
Wrong penalty (\texttt{WRONG\_PENALTY}) & -1.5 \\
Cost of S (\texttt{COST\_S}) & 0.0 \\
Cost of R (\texttt{COST\_R}) & 0.1 \\
Waste penalty for unnecessary R (\texttt{WASTE\_R}) & 0.4 \\
Format error penalty (\texttt{FORMAT\_ERROR\_PENALTY}) & -1.0 \\
\bottomrule
\end{tabular}
\caption{Reward coefficients used by the router GRPO objective.}
\end{table}


\section{Detailed Analysis of Provenance Pointer Ablation}
\label{app:provenance}

This appendix provides a detailed analysis of the provenance pointer ablation study introduced in Section~\ref{sec:ablation_provenance}.
We analyze route consistency, accuracy under identical routing decisions, and token-level cost, in order to isolate the effect of provenance-aware warm-start from routing behavior or additional computation.

\subsection{Route Consistency}
\label{app:provenance_route}
Across 1,540 shared queries, Linked and No-Linked make identical routing decisions on 1,306 queries (84.8\%), and differ on 234 queries (15.2\%).
This indicates that enabling provenance pointers induces only modest changes in routing behavior.
Therefore, improvements observed in the Linked setting cannot be explained solely by more aggressive escalation, but instead suggest higher-quality evidence retrieval during escalation.

\subsection{Accuracy under Identical Routes}
\label{app:provenance_same_route}
To disentangle retrieval quality from routing effects, we analyze queries where Linked and No-Linked select the same route.

\paragraph{Both Escalate (R).}
When both systems escalate to raw logs, Linked answers 31 queries correctly that No-Linked answers incorrectly, whereas the reverse occurs in 25 cases.
This yields a net gain of +6 queries for Linked under identical escalation decisions.

\paragraph{Both Answer (S).}
When both systems answer directly from summaries, Linked answers 16 queries correctly that No-Linked answers incorrectly, compared to 10 in the opposite direction.
The small difference confirms that provenance pointers primarily affect raw evidence grounding rather than summary-only answering.

These controlled comparisons demonstrate that provenance pointers improve the \emph{quality} of retrieved evidence, independent of routing frequency.

\subsection{Cost Analysis by Route}
\label{app:provenance_cost}
Table~\ref{tab:provenance_cost} reports average token usage by route.
On summary-only queries, Linked and No-Linked incur nearly identical cost.
On escalated queries, Linked incurs slightly higher token usage, reflecting deeper or more precise raw evidence grounding.

Importantly, the average research depth is nearly identical between Linked (1.38) and No-Linked (1.36), with the same maximum depth (3), indicating that gains do not stem from additional retrieval iterations.

\begin{table}[H]
\centering
\small
\setlength{\tabcolsep}{6pt}
\begin{tabular}{lccc}
\toprule
\textbf{Method} & \textbf{Route} & \textbf{Tok$_{\text{in}}$} & \textbf{Tok$_{\text{out}}$} \\
\midrule
Linked & S & 847.5 & 97.8 \\
Linked & R & 8875.1 & 724.0 \\
\midrule
No-Linked & S & 847.2 & 96.8 \\
No-Linked & R & 8705.0 & 709.9 \\
\bottomrule
\end{tabular}
\caption{Average QA token cost by route for provenance pointer ablation.}
\label{tab:provenance_cost}
\end{table}

\subsection{Summary}
\label{app:provenance_summary}
In summary, provenance pointers consistently improve performance on escalated queries by enabling more reliable raw-log grounding.
The improvement persists under identical routing decisions and does not rely on increased search depth.
While provenance-aware escalation incurs slightly higher token cost, it yields a favorable accuracy--efficiency trade-off, validating provenance linking as a core component of TierMem.



\section{Additional Analysis: Why Higher Hard-Recall Can Tie on End-to-End Judge Score}
\label{app:router_counterfactual}

This appendix explains why GPT-4.1-mini achieves higher recall on oracle-hard cases (Table~\ref{tab:router_learning}),
yet matches our SFT+GRPO router on the final LoCoMo LLM-as-judge accuracy.
Crucially, this effect does \emph{not} come from different stored memories:
all variants share identical writes and use the same memory library, generator, embedding model, and reranker (cf.\ evaluation protocol in \S\ref{sec:baselines}).
Instead, the tie arises from how routing interacts with (i) the \emph{difficulty distribution} of escalated queries and
(ii) the \emph{effective repair/regression behavior} of the R-path under the realized evidence trajectories.

\subsection{S/R counterfactual comparison matrices on the R-routed set}
\label{app:router_counterfactual_matrices}

For each router, we log a counterfactual comparison on the subset of queries routed to R:
we run the S-path answer and the R-path answer and record whether each is correct under the benchmark judge.
This yields a 2$\times$2 table with counts:
\begin{itemize}
    \item \textbf{S\_correct\_R\_correct}: both paths would be correct.
    \item \textbf{S\_correct\_R\_wrong}: R would \emph{hurt} (regression if escalated).
    \item \textbf{S\_wrong\_R\_correct}: R would \emph{fix} S (repair benefit).
    \item \textbf{S\_wrong\_R\_wrong}: both fail (hard negatives).
\end{itemize}
We report both (i) the raw S/R agreement counts and (ii) their decomposition against the final judge labels (\texttt{actual\_correct/wrong})
to expose how ``hard'' the routed set is.

\paragraph{GPT-4.1-mini (R-routed subset).}
\begin{table}[t]
\centering
\small
\setlength{\tabcolsep}{6pt}
\begin{tabular}{lrr}
\toprule
\textbf{Category} & \textbf{Count} & \textbf{Decomposition (actual)} \\
\midrule
S\_correct\_R\_correct & 240 &
225 correct / 15 wrong \\
S\_correct\_R\_wrong & 33 &
13 correct / 20 wrong \\
S\_wrong\_R\_correct & 181 &
169 correct / 12 wrong \\
S\_wrong\_R\_wrong & 86 &
14 correct / 72 wrong \\
\midrule
Total matched (R set) & 540 & 421 correct / 119 wrong \\
\bottomrule
\end{tabular}
\caption{GPT-4.1-mini: S/R counterfactual comparison on the queries routed to R.}
\label{tab:gpt41_sr_matrix}
\end{table}

\paragraph{Ours (SFT+GRPO) (R-routed subset).}
\begin{table}[t]
\centering
\small
\setlength{\tabcolsep}{6pt}
\begin{tabular}{lrr}
\toprule
\textbf{Category} & \textbf{Count} & \textbf{Decomposition (actual)} \\
\midrule
S\_correct\_R\_correct & 313 &
305 correct / 8 wrong \\
S\_correct\_R\_wrong & 26 &
13 correct / 13 wrong \\
S\_wrong\_R\_correct & 170 &
153 correct / 17 wrong \\
S\_wrong\_R\_wrong & 92 &
20 correct / 72 wrong \\
\midrule
Total matched (R set) & 601 & 491 correct / 110 wrong \\
\bottomrule
\end{tabular}
\caption{Ours (SFT+GRPO): S/R counterfactual comparison on the queries routed to R.}
\label{tab:ours_sr_matrix}
\end{table}

\subsection{Repair benefit vs.\ regression risk}
\label{app:router_repair_regress}

Hard-recall is a mechanism metric that measures whether oracle-hard cases are escalated,
but end-to-end judge accuracy depends on whether escalation \emph{actually repairs} S failures without introducing regressions.
From Tables~\ref{tab:gpt41_sr_matrix}--\ref{tab:ours_sr_matrix}, we define two diagnostics on the R-routed subset:
\begin{align}
\mathrm{RepairRate} &= \Pr(R\ \text{correct} \mid S\ \text{wrong})
= \frac{\text{S\_wrong\_R\_correct}}{\text{S\_wrong\_R\_correct}+\text{S\_wrong\_R\_wrong}}, \\
\mathrm{RegressRate} &= \Pr(R\ \text{wrong} \mid S\ \text{correct})
= \frac{\text{S\_correct\_R\_wrong}}{\text{S\_correct\_R\_wrong}+\text{S\_correct\_R\_correct}}.
\end{align}

\paragraph{Measured values.}
For GPT-4.1-mini: $\mathrm{RepairRate}=181/(181+86)=67.8\%$ and $\mathrm{RegressRate}=33/(33+240)=12.1\%$.
For ours: $\mathrm{RepairRate}=170/(170+92)=64.9\%$ and $\mathrm{RegressRate}=26/(26+313)=7.7\%$.
Thus, GPT-4.1-mini shows higher repair rate but also higher regression exposure on the set it escalates,
whereas our router is slightly less aggressive in ``fixing'' but produces fewer cases where R would be worse than S.

\subsection{Why end-to-end accuracy can tie despite recall differences}
\label{app:router_tie_explanation}

Even with identical writes and identical components, routers induce different \emph{escalated sets} and therefore different
difficulty distributions for the R-path.
This is visible in the \textbf{hard-negative} bucket \texttt{S\_wrong\_R\_wrong}:
GPT-4.1-mini escalates 86 such cases, while ours escalates 92.
Importantly, within this bucket the judge decomposition shows that some instances labeled as \texttt{S\_wrong\_R\_wrong}
still end up as \texttt{actual\_correct} (14 for GPT-4.1-mini vs.\ 20 for ours), which we interpret as evidence of
\emph{trajectory sensitivity} in the R-path: small differences in retrieved evidence traces and generation can flip outcomes on extremely difficult instances.
In our runs, the R-path succeeds on more of these hard negatives (20 vs.\ 14), partially compensating for the lower oracle-hard recall.

\paragraph{Takeaway.}
Under a fixed write state, the gap between mechanism metrics (e.g., oracle-hard recall) and outcome metrics (judge accuracy)
can be explained by (i) different escalated-set difficulty, (ii) repair vs.\ regression trade-offs, and
(iii) trajectory sensitivity of the R-path on borderline hard instances.
Therefore, we report not only hard recall but also the S/R counterfactual matrices to make the cost--fidelity behavior transparent.

\section{Router Error Analysis and Failure Modes}
\label{sec:router_error_analysis}
We analyze routing errors relative, clustered into coherent categories (Table~\ref{tab:router_error_analysis}).
We distinguish two asymmetric error types:
\textbf{(i) S$\rightarrow$R errors} (false cache hits), where the router answers from Tier-1 but raw grounding is required; and
\textbf{(ii) R$\rightarrow$S errors} (false cache misses), where the router escalates unnecessarily despite sufficient summary evidence.

\paragraph{False cache hits are rarer but more harmful.}
S$\rightarrow$R errors are relatively rare (61 cases) but directly impact correctness.
They stem mainly from \emph{information completeness} failures (missing a required slot such as an entity name) and \emph{precision requirement} failures (ignoring that the question demands exactness or attribution).
These cases align with known brittleness of lossy compression: a summary can be topically relevant yet underspecified for exact answering.

\paragraph{False cache misses dominate and reflect conservative bias.}
R$\rightarrow$S errors are much more frequent (371 cases).
The primary cause is \emph{false information gap detection}: the router assumes summaries are insufficient even when they contain enough evidence.
These errors increase cost but typically do not degrade correctness, reflecting a conservative policy that prioritizes faithfulness.
This suggests a clear path for future improvement: better calibration and sufficiency detection that leverages indirect but adequate evidence while still avoiding false cache hits.

\begin{table*}[!t]
\centering
\small
\setlength{\tabcolsep}{6pt}
\begin{tabular}{l r}
\toprule
\textbf{Error Category} & \textbf{Count} \\
\midrule
\multicolumn{2}{l}{\textbf{S$\rightarrow$R Errors (Summary insufficient, escalation needed)}} \\
\midrule
Information completeness error & 26 \\
Disambiguation failure & 17 \\
Precision requirement ignored & 16 \\
Context and focus confusion & 1 \\
Other & 1 \\
\midrule
\textbf{Total S$\rightarrow$R} & \textbf{61} \\
\midrule\midrule
\multicolumn{2}{l}{\textbf{R$\rightarrow$S Errors (Unnecessary escalation)}} \\
\midrule
False information gap detection & 289 \\
Neglect of indirect or inferential evidence & 49 \\
Temporal or logical inference failure & 33 \\
\midrule
\textbf{Total R$\rightarrow$S} & \textbf{371} \\
\bottomrule
\end{tabular}
\caption{\textbf{Router error analysis.}
We categorize routing errors into false cache hits (S$\rightarrow$R) and false cache misses (R$\rightarrow$S).
False cache hits are relatively rare but stem mainly from summary incompleteness and unmet precision requirements.
False cache misses dominate and primarily reflect conservative over-escalation, incurring additional cost without affecting correctness.}
\label{tab:router_error_analysis}
\end{table*}

\section{Prompts Used in the TierMem System}
\label{app:prompts}

\begin{figure*}[htb]
\begin{AIbox}[width=\textwidth]{Fact Extraction Prompt (JSON Schema)}
\footnotesize
You are the \textbf{Universal Memory Encoder}. Your goal is to convert raw input stream into high-fidelity, self-contained knowledge records (Long-term Memory).

INPUT FORMATTING NOTICE: \newline
The input will be provided in the next user message, prefixed by `Input:' and often formatted as dialogue lines. \newline
It may follow a pattern like: `[Timestamp] Speaker\_Name: Content'.
\begin{itemize}
    \item If this pattern is present, you MUST use the `Timestamp' for temporal grounding and `Speaker\_Name' for entity resolution.
    \item If this pattern is absent (e.g., raw document text), treat the input as a factual source and extract knowledge propositions.
\end{itemize}

\textbf{CORE OBJECTIVES:}
\begin{enumerate}
    \item \textbf{ENTITY \& CONTEXT RESOLUTION}:
    \begin{itemize}
        \item \textbf{No ambiguous references}: Every extracted fact must explicitly name the involved people, organizations, objects, places, and concepts.
        \item \textbf{No ``User'' ambiguity}: If the input says `[... ] Melanie: I like art', the fact MUST be ``Melanie likes art'', NOT ``The user likes art''.
        \item \textbf{De-contextualization}: Each extracted fact must be \textbf{standalone}.
    \end{itemize}
    \item \textbf{TEMPORAL GROUNDING}: Use the provided timestamps in the input as the source of truth. Convert relative time (``tomorrow'', ``next week'') into absolute context.
    \item \textbf{PRONOUN \& DEIXIS ELIMINATION}: Extracted facts must \textbf{not contain any pronouns}. If a pronoun appears, resolve it to the specific entity.
    \item \textbf{SCENE TAGGING (EMBEDDED)}: Identify the immediate scene/activity (2-5 words). Append this scene to the end of the fact string inside brackets.
\end{enumerate}

\textbf{OUTPUT INSTRUCTIONS (MUST FOLLOW):} \newline
Return a JSON object with EXACTLY this schema:
\begin{verbatim}
{
  "facts": [
      "<atomic fact sentence 1> [Scene: ...]", 
      "<atomic fact sentence 2> [Scene: ...]", ...
  ]
}
\end{verbatim}
Rules: Output JSON only (no extra text, no markdown code fences).
\end{AIbox}
\caption{The prompt used for extracting atomic facts from conversation streams.}
\label{fig:fact_extraction_prompt}
\end{figure*}

\begin{figure*}[htb]
\begin{AIbox}[width=\textwidth]{Research Integration Prompt}
\footnotesize
You are a research assistant extracting facts from conversation evidence to answer a question.

\textbf{QUESTION:} \{question\} \newline
\textbf{RETRIEVED EVIDENCE:} \{evidence\}

\textbf{YOUR TASK:} \newline
Read through ALL pages in the evidence and extract facts that help answer the question.

\textbf{EXTRACTION STRATEGY:} \newline
Identify what information is needed and extract:
\begin{enumerate}
    \item \textbf{Direct answers}: Facts that directly answer the question
    \item \textbf{Component facts}: Facts about entities/topics in the question that can be combined
    \item \textbf{Temporal facts}: When events happened
    \item \textbf{Confirmation facts}: Look for acceptance/approval messages
\end{enumerate}

\textbf{OUTPUT FORMAT (JSON):}
\begin{verbatim}
{
  "linked_facts": [
    {
      "fact": "<extracted fact - use specific dates/names>",
      "evidence_quote": "<EXACT quote from the evidence>"
    }
  ],
  "coverage_assessment": "<what aspects are covered vs missing>"
}
\end{verbatim}
\end{AIbox}
\caption{Prompt for integrating retrieved facts to answer a query.}
\label{fig:research_integration_prompt}
\end{figure*}

\begin{figure*}[htb]
\begin{AIbox}[width=\textwidth]{Research Plan Prompt}
\footnotesize
You are a research planner. Based on the current facts and the question, decide if more information is needed.

\textbf{QUESTION:} \{question\} \newline
\textbf{CURRENT FACTS:} \{current\_facts\} \newline
\textbf{COVERAGE ASSESSMENT:} \{coverage\_assessment\} \newline
\textbf{RESEARCH HISTORY:} \{research\_history\} \newline
\textbf{ALREADY SEARCHED:} \{searched\_queries\}

\textbf{YOUR TASK:} \newline
Decide whether the current facts are SUFFICIENT to answer the question, or if more search is needed.

\textbf{OUTPUT FORMAT (JSON):}
\begin{verbatim}
{
  "decision": "DONE" or "SEARCH",
  "reasoning": "<brief explanation>",
  "search_commands": [
    {"type": "MEM0_SEARCH", "query": "<semantic search query>"},
    {"type": "KEYWORD_SEARCH", "keywords": ["k1", "k2"]}
  ]
}
\end{verbatim}

\textbf{DECISION CRITERIA:}
\begin{itemize}
    \item \textbf{DONE}: The facts contain enough information to provide a reasonable answer.
    \item \textbf{SEARCH}: Key information is missing and more search might help.
\end{itemize}
\end{AIbox}
\caption{Prompt for planning the next research step or concluding the search.}
\label{fig:research_plan_prompt}
\end{figure*}

\begin{figure*}[htb]
\begin{AIbox}[width=\textwidth]{Router Prompt (Summarization vs. Research Decision)}
\footnotesize
You are an expert router for a memory-augmented QA system. Analyze the retrieved summaries and decide the best action to answer the question.

\textbf{Available actions:} \newline
1. "S" - Answer using current summaries only. Use when Summaries contain the EXPLICIT answer. \newline
2. "R" - Deep research mode (slow path). Use when Summaries are ambiguous, only contextually related, or miss the answer entirely.

Question: \{q\} \newline
Retrieved Summaries: \{summaries\_block\}

Output format (JSON only):
\begin{verbatim}
- If answering with summaries: {"action": "S"}
- If deep research needed: {"action": "R"}
\end{verbatim}
\end{AIbox}
\caption{Routing prompt to decide between immediate summarization or deep research.}
\label{fig:router_prompt}
\end{figure*}

\begin{figure*}[htb]
\begin{AIbox}[width=\textwidth]{Answer Generation Prompts (Template Functions)}
\footnotesize
\begin{verbatim}
def _make_locomo_summary_prompt(summary: str, question: str) -> str:
    return f"""\
Based on the summary below, write an answer in the form of **a short phrase**...
Answer with exact words from the context whenever possible.
For date/time, strictly follow the format "15 July 2023"...

QUESTION: {question}
SUMMARY: {summary}
Short answer:
"""

def _make_longmemeval_prompt(summary: str, question: str) -> str:
    return f"""\
You are an intelligent memory assistant tasked with retrieving accurate 
information from episodic memories.

# INSTRUCTIONS:
Synthesize information from different memories to answer the user's question.
It is CRITICAL that you move beyond simple fact extraction and perform 
logical inference. Answer the question in a short phrase.

QUESTION: {question}
SUMMARY: {summary}
Short answer:
"""
\end{verbatim}
\end{AIbox}
\caption{Python functions used to generate the final answer prompts based on context.}
\label{fig:answer_prompts}
\end{figure*}

\begin{figure*}[htb]
\begin{AIbox}[width=\textwidth]{Write-Back Grounding Verification Prompt}
\footnotesize
You are a strict quality judge for memory facts. Given a question and candidate facts, select ONLY high-quality facts that meet ALL criteria.

Question: \{question\} \newline
Candidate Facts: \{facts\_list\}

\textbf{Quality Criteria (ALL must be met):}
\begin{enumerate}
    \item \textbf{Directly Relevant}: Directly helps answer this specific question.
    \item \textbf{Specific \& Concrete}: Contains names, dates, numbers, locations.
    \item \textbf{Factually Grounded}: Based on concrete conversation content.
    \item \textbf{Non-redundant} \& \textbf{Self-contained}.
\end{enumerate}

\textbf{Reject facts that are:} Too general, obvious from the question, vague, or tangentially related.

\textbf{Output format:} Return ONLY a JSON array of selected fact indices (0-based), e.g., [0, 2], [1], or [] if none qualify.
\end{AIbox}
\caption{Prompt for verifying the quality and relevance of extracted facts.}
\label{fig:write_back_prompt}
\end{figure*}

\begin{figure*}[htb]
\begin{AIbox}[width=\textwidth]{LLM as Judge Evaluation Prompt}
\footnotesize

Your task is to label an answer to a question as `CORRECT' or `WRONG'. 
You will be given:
    (1) a question (posed by one user to another user), 
    (2) a `gold' (ground truth) answer, 
    (3) a generated answer.

The gold answer will usually be a concise and short answer. 
The generated answer might be much longer, but you should be generous 
with your grading - as long as it touches on the same topic/time 
as the gold answer, it should be counted as CORRECT.

Question: \{question\}
Gold answer: \{gold\_answer\}
Generated answer: \{generated\_answer\}

First, provide a short (one sentence) explanation of your reasoning, 
then finish with CORRECT or WRONG in a json format with the key as "label".

\end{AIbox}
\caption{The LLM-as-a-Judge prompt used for automated evaluation.}
\label{fig:llm_judge_prompt}
\end{figure*}

\end{document}